\documentclass{article}

\usepackage{arxiv}

\usepackage[utf8]{inputenc} 
\usepackage[T1]{fontenc}    
\usepackage{url}            
\usepackage{booktabs}       
\usepackage{amsfonts}       
\usepackage{nicefrac}       
\usepackage{orcidlink}
\usepackage{multirow}
\usepackage{microtype}      
\usepackage{lipsum}
\usepackage{graphicx}
\usepackage{bm}
\usetikzlibrary{trees}

\graphicspath{ {./images/} }
\usepackage{tikz-qtree}
\usepackage{tikz}
\usepackage{tikz}
\usetikzlibrary{shapes.geometric, arrows}
\tikzstyle{startstop} = [rectangle, rounded corners, minimum width=3cm, minimum height=1cm,text centered, draw=black, fill=red!30]
\tikzstyle{process} = [rectangle, minimum width=3cm, minimum height=1cm, text centered, draw=black, fill=blue!30]
\tikzstyle{decision} = [diamond, minimum width=3cm, minimum height=1cm, text centered, draw=black, fill=green!30]
\tikzstyle{arrow} = [thick,->,>=stealth]

\RequirePackage{amsthm,amsmath,amsfonts,amssymb}

\RequirePackage{algorithm}
\RequirePackage[noend]{algpseudocode}

\RequirePackage[numbers]{natbib}
\RequirePackage{fontenc,inputenc,subcaption,graphicx,sectsty,titlesec,caption}
\RequirePackage{amsthm,amsmath,amssymb,animate,nccmath}

\RequirePackage{paralist}
\algnewcommand\REQUIRE{\State \textbf{Require:}}
\algnewcommand\ENSURE{\State \textbf{Ensure:}}
\algnewcommand\RETURN{\State \textbf{return}}

\theoremstyle{remark}

\def\b1{\mbox{$\mathbf{1}$}}

\def\bs{\mbox{\boldmath$s$}}

\def\bx{\mbox{$\mathbf{x}$}}
\def\bX{\mbox{$\mathbf{X}$}}

\def\E{\mbox{$\mathbb{E}$}}

\def\bbeta{\mbox{\boldmath$\beta$}}

\newcommand{\bvec}{\mathbf{b}}

\newcommand{\xvec}{\mathbf{x}}

\newcommand{\Xvec}{\mathbf{X}}
\newcommand{\Yvec}{\mathbf{Y}}

\newcommand{\uvec}{\mathbf{u}}

\newcommand{\epsilonvec}{\boldsymbol{\epsilon}}

\newcommand{\thetavec}{\boldsymbol{\theta}}

\newcommand{\betavec}{\boldsymbol{\beta}}

\newcommand{\diag}{\text{diag}}
\def\argmin{\operatorname*{arg\,min}}

\theoremstyle{plain}

\theoremstyle{definition}

\theoremstyle{remark}

\def\b1{\mathbf{1}}

\def\bs{\boldsymbol{s}}

\def\bx{\mathbf{x}}
\def\bX{\mathbf{X}}

\def\E{\mathbb{E}}

\def\bbeta{\boldsymbol{\beta}}

\title{Interval Estimation of Coefficients in Penalized Regression Models of Insurance Data }

\author{
Alokesh Manna \orcidlink{0009-0007-7958-5273} \\
Department of Statistics \\
University of Connecticut, Storrs, USA \\
\texttt{alokesh.manna@uconn.edu} \\
\And
Zijian Huang \\
Wells Fargo \\
Arlington, VA, USA\\
\texttt{yellowzijian@gmail.com} \\
\And
Dipak K. Dey \\
Department of Statistics \\
University of Connecticut, Storrs, USA \\
\texttt{dipak.dey@uconn.edu} \\
\And
Yuwen Gu \\
Department of Statistics \\
University of Connecticut, Storrs, USA \\
\texttt{yuwen.gu@uconn.edu} \\
\And
Jichao He \\
The Travelers Companies, Inc. \\
Hartford, CT, USA \\
\texttt{jhe@travelers.com}
}

\begin{document}
\maketitle
\begin{abstract}
The Tweedie exponential dispersion family is a popular choice to model insurance losses that consist of zero-inflated semi continuous data. In such data, it is often important to obtain credible inference on the most important features that describe the endogenous variables. Post-selection inference is the standard procedure in statistics to obtain confidence intervals of model parameters after performing a feature selection step. For a linear model, the lasso estimate often exhibits non-negligible estimation bias for large coefficients corresponding to exogenous variables. To achieve valid inference on those coefficients, it is necessary to correct this bias. Traditional statistical methods—such as hypothesis testing or standard confidence interval construction—may lead to incorrect conclusions during post-selection, as they are generally too optimistic. Here we discuss several methodologies for constructing confidence intervals of the coefficients after feature selection in the Generalized Linear Model (GLM) family, with application to insurance data.

\end{abstract}

\section{Introduction} Feature selection is a pivotal step in building efficient, accurate, and interpretable predictive models in the insurance industry. It enhances model performance, reduces computational cost, simplifies the model structure, and ensures compliance with regulatory standards. By focusing on the most relevant features, insurance companies can make better-informed decisions, improve risk assessment, and ultimately provide better services to their customers.

Consider an insurance company analyzing historical claim data to estimate the average cost of claims for a particular type of policy. Instead of providing a single point estimate (e.g., the average claim cost is \$5,000), the company calculates a 95\% confidence interval for this estimate (e.g., the average claim cost is between \$4,500 and \$5,500). This interval indicates that there is a 95\% chance that the true average claim cost lies within this range. Knowing the range of the true effect of a variable, helps in assessing the potential impact of changes in that variable. Confidence intervals allow for scenario analysis by providing a range of potential outcomes. This helps in planning for best-case, worst-case, and most likely scenarios, leading to more robust risk management strategies.

A key difficulty arises when the number of variables (features) is much larger than the number of available samples. This is common in fields like insurance, where the number of potential features (such as customer attributes) may be high, but the number of customers (or observations) might still be moderate. For example, you may have thousands of customer attributes but only a few hundred customer records.

To handle such cases, sparsity assumptions are often used, which means that only a small number of variables truly influence the outcome, while the rest are either irrelevant or have minimal effects. There are many available procedures to construct confidence intervals after the
post-selection of important variables. Sparsity assumption is one major area of
research when the number of variables is significantly higher than the number of
available samples. For insurance data, the number of available features is
generally moderate, but the customer database is very large. The key challenge
in these models is that the inference has to be drawn in a generalized linear
model setup. When using lasso (a popular method for variable selection), one challenge is that the process of selecting variables introduces model uncertainty. The lasso method automatically selects the most influential variables, but this selection process itself is part of the model. As a result, standard methods of statistical inference, like p-values or confidence intervals, may not be accurate because they do not account for the fact that the model was selected after seeing the data.

In simple terms, when you choose which variables to include in your model based on the data, this choice introduces a form of bias. Therefore, traditional methods that rely on the assumption that the model is fixed and not selected after looking at the data might not be valid anymore. This phenomenon
is known as selective inference or post-selection inference. Once the model is
selected, inference is carried out on the selected variables. This involves
adjusting the standard errors, p-values, or confidence intervals to account for
the selection process. Using the adjusted statistics, it is possible to obtain
valid conclusions about the model parameters. Conditioning on the selected model
with a small set of predictors and testing over all possible data that would
lead to this exact set of predictors being selected is a well-known two-step
procedure for conditional testing in the literature
\citep[see][]{benjamini2010simultaneous,zhang2022post}. Standard procedures such
as de-biased estimates of the coefficients are available in the literature; see
for example, \cite{zhang2014confidence} and \cite{cai2023statistical}. Some
bootstrap-related procedures are also available, for example,
\cite{chatterjee2010asymptotic}. Many of the existing procedures are based on
the normality assumption of the model errors.

The length of the confidence interval is also an important factor, as a shorter
interval indicates less variability and more confidence that the estimated
parameter is close to the true population parameter. \cite{liu2020bootstrap}
describe a procedure called bootstrap lasso and partial ridge to obtain shorter
confidence intervals than those of the penalized methods, regardless of whether
or not the linear models are misspecified. The number and amount of claims are important indices for making premiums for Insurance companies. In this paper, we will extend the idea of \cite{liu2020bootstrap} for different GLM i.e. the Poisson regression, Negative binomial regression, Tweedie regression, e.t.c. We will estimate the confidence intervals of
coefficients in our simulation for Poisson and negative binomial regression in a
sparse setup, where 75\% of the coefficients are zero with a good selection
rate. Another important flexible family of probability distributions is the Tweedie distribution which can
model data with different characteristics, such as count data, positive
continuous data, and data with a mixture of zeros and positive continuous
values. It is part of the exponential dispersion
family and is particularly useful in generalized linear models (GLM) when
modeling data that are not well-represented by common distributions like normal,
binomial, or Poisson distributions. Tweedie distribution is very commonly used
for modeling compound Poisson distribution. \cite{halder2021spatial} considered
a double generalized linear model (DGLM) in the Tweedie framework allowing
dispersion to depend in a link-linear fashion on chosen features with an
application for modeling insurance losses arising from automobile collisions in
the state of Connecticut, USA for the year 2008.

There are several Bayesian perspectives on Tweedie models which has been
explored in the literature. \cite{halder2023bayesian} discussed Bayesian
variable selection or feature extraction in the double GLM
framework for Tweedie spatial process modeling. \cite{ye2021comparisons}
discussed a Bayesian perspective of Zero Augmented gamma model and Tweedie model
with sensitivity analysis of the different choice of priors for the power
parameter $p$ such as uniform, beta distribution, e.t.c.

First, we provide a brief overview how feature selection has been a key discussion topic in insurance data analysis in section\ref{literature_insurance_feature_selection}. We introduce a GLM in section \ref{glm}. Next, we
explain the Tweedie distribution in section \ref{tweedie} and standard variable
selection procedures for GLM in section \ref{pr}. Then we illustrate a few
available methodologies for confidence interval estimation in penalized
regression setup in section \ref{ci_estimation} which includes de-biased
estimation for the linear model and GLM, bootstrap
estimates for the linear model and GLM with different types
of residuals such as Pearson residuals, deviance residuals and Anscombe
residuals. We have used 95\% quantile for confidence interval estimation when there is a normality assumption. The readers can extend to any $\alpha\%$ quantile. We also explain how different machine learning approaches such as
LightGBM(see \cite{ke2017lightgbm}) can also estimate the importance of
different features in the context of insurance data. In a Bayesian framework,
the confidence interval is denoted as credible interval. We also explain a few
Bayesian perspectives for credible interval. Then we explain another methodology
that combines bootstrap procedure with lasso and ridge penalization in section
\ref{plr_lm}. Then we explain penalization in Tweedie regression in section
\ref{p_tweedie}. We extend the idea of section \ref{plr_lm} for GLM in section
\ref{plr_glm}. Finally, we do apply \ref{plr_glm} with simulation results for
Poisson and Negative binomial regression setup with a real data example for
Tweedie regression for estimating confidence interval for different variables in
section \ref{simulation}. We apply the proposed methodology with Tweedie
regression for estimating the confidence intervals of the coefficients of an
insurance auto-claim dataset (\textbf{AutoClaim} data from the \texttt{cplm} R
package, see \cite{zhang2013cplm}).

\subsection{Literature survey on feature extraction in insurance data}
\label{literature_insurance_feature_selection}
Feature extraction in insurance data is a well-discussed topic in literature. In recent times, the effect of weather specifically in coastal areas has started to be explored for premium housing insurance. \cite{scheel2013bayesian} discussed a Bayesian Poison hurdle model with different covariates, for example, precipitation, temperature, drainage, snow-water equivalent and several other derived features in GLM framework. Vehicle insurance has always been a major business all over the world. \cite{baecke2017value} discussed how the sensor telematics data can improve the premium calculation of a customer with random forest and neural networks. It is very common to obtain additional coverage for specific events that might damage the insured vehicle including fire, natural disasters, theft, windscreen repair, and legal expenses. \cite{gomez2021priori} explored their methods in a multivariate regression setup on the portfolio of customers who have five different coverages in auto insurance.\cite{karadaug2019selection} explored non-life insurance-specific regression models to predict net profit and asset ratio based on premium, total assets, company size, market share, and several other important meaningful variables with fixed and random effects in a linear model setup based on 30 non-life insurance companies between 2010 and 2014 in Turkey. \cite{waters1999measuring} discussed different regression models to demonstrate the importance of different exogenous and endogenous variables to model the demand of health care. \cite{cho2003data} explored different machine learning and statistical models for the selection of insurance sales agents in the growing Hong Kong area based on their individual information available. \cite{quan2018predictive} explored the importance of decision trees for multi-line insurance coverage prediction, which includes property, motor vehicle, and contractors’ equipment for Wisconsin Local Government Property Insurance Fund (LGPIF). \cite{mladenovic2020identification} discussed different models for the identification of different variables for the prediction of individual medical costs billed by health insurance. They considered different individual specific variables such as age of primary beneficiary, insurance contractor gender, Body mass index, Number of children
covered by health insurance, and smoking and found that smoking is a key factor for an increasing cost of health insurance using adaptive neuro-fuzzy inference system (ANFIS). \cite{segovia2015risk} discussed methods to increase the prediction capacity of
‘bonus-malus’ (BM) level, and psychological questionnaires could be used to measure
policyholders’ hidden characteristics. Feature extraction is always popular even other than insurance data, such as selection of features in environment and insurance data (see~\cite{manna2024development}, \cite{manna2024interval}, e.t.c. ). Thus in the literature, we can see that feature selection is a very well-known topic in different types of insurance setups. We will focus on a well-known feature selection method using penalized regression and a proper interval estimation method to obtain an accurate confidence interval of the important coefficients.
\subsection{Introduction to GLM }\label{glm}
GLM is a rich class of probability
distributions, including many commonly used ones such as the Gaussian, Poisson,
binomial and gamma distributions. The general form of this class of
Distributions can be expressed as
\begin{equation}
\begin{aligned}
  f_{Y}(y;\theta,\phi)
  =\exp\left\{\frac{y\theta-b(\theta)}{a(\phi)}+c(y,\phi)\right\},
\end{aligned}
\end{equation}
where $\theta$ is the canonical parameter and $\phi$ is the dispersion
parameter. The mean of \(Y\) can be expressed respectively as $\mu=E(Y)=b^{'}(\theta)$, hence $\theta=(b^{'})^{-1}(E(Y))$. The variance can be also expressed as $\sigma^{2}=var(Y)=b^{''}(\theta)a(\phi)=b^{''}\{(b^{'})^{-1}(E(Y))\}a(\phi)$. The last equality shows the relationship between the variance and the mean
of a distribution from the GLM family. GLM are formulated by \cite{nelder1972_gener_linea_model}
as a way of unifying various statistical models, including linear regression, logistic regression, Poisson regression, and many others. In a GLM, we have the output variable \(Y\) from the GLM
family, given a set of features $\xvec=(x_{1},\cdots,x_{p})^{T}$. To describe
the dependence of \(Y\) on \(\xvec\), we use a link function \(g\) that links
the mean of \(Y\) to a linear function of the features:
$g(\mu)=\xvec^{T}\bbeta=\eta$. For a specific response distribution, various
link functions may be used. The canonical link $g=(b^{'})^{-1}$ is commonly used
in practice, for example: linear regression: $\eta=\mu$, Binomial regression: $\eta=\log(\frac{\mu}{1-\mu})$, where $\mu=E(Y/m)$, Poisson regression: $\eta=\log(\mu)$, Gamma regression: $\eta=-\frac{1}{\mu}$ are some commonly used models in practice. 
The iteratively re-weighted least squares (IRLS) method is typically used to compute the maximum likelihood estimation of the model parameters. In this case the likelihood function for independent and identically distributed (i.i.d.) observations
$\{y_{i},\bx_{i}\}_{i=1}^{n}$ and $a_{i}(\phi)=\phi/w_{i}$ can be written as 
\[L(\bbeta,\phi)=\prod_{i=1}^{n}\exp\biggl\{\frac{w_{i}}{\phi}
      \Bigl[y_{i}h(\bx_{i}^{T}\bbeta)-b\bigl(h(\bx_{i}^{T}\bbeta)\bigr)
      \Bigr]+c(y_{i},\phi)\biggr\}.
\] Hence the log-likelihood can be expressed as
\[
-\frac1n\log{}L(\bbeta,\phi)
      =-\frac{1}{n}\sum_{i=1}^{n}\biggl\{\frac{w_{i}}{\phi}
      \Bigl[y_{i}h(\bx_{i}^{T}\bbeta)-b\bigl(h(\bx_{i}^{T}\bbeta)\bigr)
      \Bigr]+c(y_{i},\phi)\biggr\}.
\] Here $\theta(\bx)=(b')^{-1}(\mu(\bx))=(b')^{-1}(g^{-1}(\eta(\bx)))=h(\bx^{T}\bbeta)$, so $h=(g\circ b')^{-1}$ with the canonical link function $g=(b')^{-1}$, so $\theta(\bx)=\bx^{T}\bbeta$. See \cite{nelder1972_gener_linea_model} and
\cite{mccullagh89_genera_linear_models} for more details.

\subsection{Introduction to Tweedie distribution} \label{tweedie}
Here we describe Tweedie distribution which is very popular in insurance data modeling. The model can be written as 
\[
    f(y \mid p, \mu, \phi) = f(y \mid p, y, \phi) \exp\left(-\frac{1}{2\phi} d(y, \mu)\right),
\] where $d(y, \mu)$ is the deviance which is expressed as 
\[
    d(y, \mu) = 2 \cdot \left(\frac{y^{2-p}}{(1-p) \cdot (2-p)} - \frac{y \cdot \mu^{1-p}}{1-p} + \frac{\mu^{2-p}}{2-p}\right)
\] (see \cite{dunn2008evaluation}). The Tweedie model is closely connected to the  exponential dispersion model, which is given in the form of
\[
  f_Y(y;\theta, \phi)=c(y, \phi) \exp \left( \frac{y \theta-b(\theta)}{a(\phi)} .\right)
\]
 Tweedie model has a parameter called the power value $p$, which makes a relationship with mean and variance in the Tweedie family. For different choices of $p$, one can obtain different distributions. For example $p = 0 $ implies Normal distribution, $p = 1$ Poisson distribution, $1 < p < 2$  Compound Poisson/gamma distribution,
$p = 2$ gamma distribution,
$2 < p < 3$ positive stable distributions,
$p = 3$ inverse Gaussian distribution,
$p > 3$ positive stable distributions and
$p = \infty$ extreme stable distributions can be obtained.
The compound Poisson distribution is useful in scenarios, where we need to model aggregate losses in insurance, total claims, or other forms of accumulated risk. It is also applied in various fields such as finance, telecommunications, and environmental studies.
Formally, let \( N \) be a Poisson-distributed random variable defining the number of customers of an insurance company with rate parameter \( \lambda \), so that
\[
P(N = k) = \frac{\lambda^k e^{-\lambda}}{k!}, \quad k = 0, 1, 2, \ldots
\]
Let \( X_1, X_2, \ldots, X_N \) be i.i.d. random variables with distribution \( F_X(.) \) defining the claim amount of each of the customers. The compound Poisson random variable \( Y \) is defined as \[Y = \sum_{i=1}^{N} X_i,\] where \( Y \) is the total amount of possible claims. In insurance data, typically many of the customers do not make any claim in a time interval resulting in a spike in zero for histogram density of \(Y\) and then followed by a regular exponential curve. Thus in an insurance setup, one analyst is interested in the compound Poisson model which can be obtained by Tweedie distribution, where $p$ is in between 1 to 2. So we could reform the likelihood function in the form of EDMs
\[
  \mu=b'(\theta), V(\mu)=b''(\theta)=\mu^p .%
\]
Now to obtain the function $b(\theta)$, one can do the following steps.
\[
  \frac{d \theta}{d \mu} = \mu^{-p} \Rightarrow \theta = \frac{\mu^{1-p}}{1-p},
  \frac{d b}{d \mu} = b'(\theta) \frac{d \theta}{d \mu} = \mu^{1-p} \Rightarrow
  b(\theta) = \frac{\mu^{2-p}}{2-p}. %
\]
In that case, the density can be written as
\[
  f_Y(y;\theta, \phi)=c(y,\phi) \exp\left\{ \frac{y\mu^{1-p}}{\phi(1-p)} - \frac{\mu^{2-p}}{\phi(2-p)} \right\},%
\]
where $a(\phi)=\phi$, $c(y, \phi)$ is infinite series summation \citep{peter2005}.
Here $\eta(\bx_{i})=F(\bx_{i})$ and $\mu(\bx_{i})=E(Y\mid\bx_{i})=\exp(F(\bx_{i}))=\exp(\bx_{i}^{T}\betavec)$. The Loss function i.e., negative log-likelihood is
\[\frac{1}{n}\sum_{i=1}^{n}\left\{-y_{i}\frac{\exp[(1-p)F(\bx_{i})]}{1-p}+\frac{\exp[(2-p)F(\bx_{i})]}{2-p}\right\}.\] 
\subsection{Penalized regression}\label{pr}
In many studies, we must determine which predictors are important to the
outcome variable. A common approach is shrinkage estimation, which
simultaneously estimates the effects of features and selects the predictors. Let
$\Yvec=(y_{1},y_{2},\cdots,y_{n})^{T}$ and
$\Xvec=(\bx_{1}^{T},\bx_{2}^{T},\cdots,\bx_{n}^{T})^{T},$ where $\bx_{i}^{T}=\left(x_{i1},x_{i1},\cdots,x_{ip}\right)$ is a $p$ dimensional vector for each $i\in\{1,2,\cdots,n\}$. For the linear model, we minimize the loss function with respect to $\bbeta$ for the quadratic loss
 $ \frac{1}{n}\|\Yvec-\Xvec\bbeta\|_{2}^{2}$ subject to a penalty term $P(\bbeta,\lambda)$. In general,
 \[\hat{\bbeta}_{n}(\Yvec,\Xvec,\lambda)
  =\argmin_{\bbeta}
  \frac{1}{n}\|\Yvec-\Xvec\bbeta\|_{2}^{2}+ P(\bbeta,\lambda).\]
One can regularize the residual sum of squares with different penalty terms $P(\bbeta,\lambda)$, such as for lasso $\lambda \|\bbeta\|_{1} = \lambda \sum_{j=1}^{p} |\beta_{j}|$ (see \cite{tibshirani1996regression}), for ridge $\lambda \|\bbeta\|_{2}^{2} = \lambda \sum_{j=1}^{p} \beta_{j}^{2}$ (see \cite{hoerl1970ridge}), for elastic net $\lambda_{1} \|\bbeta\|_{1} + \lambda_{2} \|\bbeta\|_{2}^{2}$ (see \cite{zou2005regularization}), for adaptive lasso $\lambda \sum_{j=1}^{p} \hat{w}_{j}|\beta_{j}|$ (see \cite{zou2006adaptive}); etc., where \(\lambda, \lambda_{1}, \lambda_{2}\) are the tuning parameters to control the amount of regularization applied to the model and \(\hat{w}_{j}\) are the weights applied to each coefficient for adaptive lasso. For all the above cases, we have
$\lambda,\lambda_{1},\lambda_{2},\hat{w}_{j}\geq0, j=1,\ldots,p$. There are a few other non-convex penalties such as smoothly clipped absolute deviation (SCAD) introduced by \cite{fan2001variable}, which attempts to remove bias in lasso while retaining the sparsity with a continuous penalization. Due to non-convexity, the optimization problem might get harder in terms of computation for GLM.
\label{poisson_nb_regression} In a penalized GLM, we need to minimize the negative log-likelihood, i.e., $-\frac{1}{n}\ell(\bbeta,\phi)$ with a penalty
term similar to the linear model case. For example, the Poisson regression can be modeled as \[\Pr(Y_{i}=y_{i}\mid\lambda_{i})=\frac{\exp(-\lambda_{i})\lambda_{i}^{y_{i}}}{y_{i}!},\]
where $\log(\lambda_{i})=\bx_{i}^{T}\bbeta$. So the negative log-likelihood is reduced to
\[ -\frac{1}{n}\ell(\bbeta,\lambda)=-\frac{1}{n}\sum_{i=1}^{n}\{-\lambda_{i}+y_{i}\log(\lambda_{i})-\log(y_{i}!)\}\\
=\frac{1}{n}\sum_{i=1}^{n}\bigl\{\exp({\bx_{i}^{T}\bbeta})
    -y_{i}\bx_{i}^{T}\bbeta+\log(y_{i}!)\bigr\}.\]
Negative Binomial Regression is a type of GLM used for modeling count data that are over-dispersed, meaning the variance exceeds the mean. It is often used as an alternative to Poisson regression when the assumption of equal mean and variance (a key assumption of the Poisson model) is violated. The probability mass function (PMF) of the Negative Binomial distribution is given by
\[
P(Y_i = y_i \mid \mu_i, \theta) = \frac{\Gamma(y_i + \theta)}{y_i! \, \Gamma(\theta)} \left( \frac{\theta}{\theta + \mu_i} \right)^\theta \left( \frac{\mu_i}{\theta + \mu_i} \right)^{y_i}, \quad y_i = 0, 1, 2, \dots,
\] where \( \Gamma(\cdot) \) is the gamma function. The mean \( \mu_i \) is linked to the linear predictor \( \mathbf{x}_i^{T} \boldsymbol{\beta} \) through a log link function $\log(\mu_i) = \mathbf{x}_i^{T} \boldsymbol{\beta}$. The negative log-likelihood function, which is minimized during the parameter estimation process, is
\[
-\frac{1}{n}\ell(\boldsymbol{\beta}, \theta) = -\frac{1}{n}\sum_{i=1}^{n} \left[ \log \Gamma(y_i + \theta) - \log \Gamma(\theta) - \log(y_i!) + \theta \log\left( \frac{\theta}{\theta + \mu_i} \right) + y_i \log\left( \frac{\mu_i}{\theta + \mu_i} \right) \right].
\]
\section{Different Methodologies for Confidence Interval Estimation for Important Features}\label{ci_estimation}
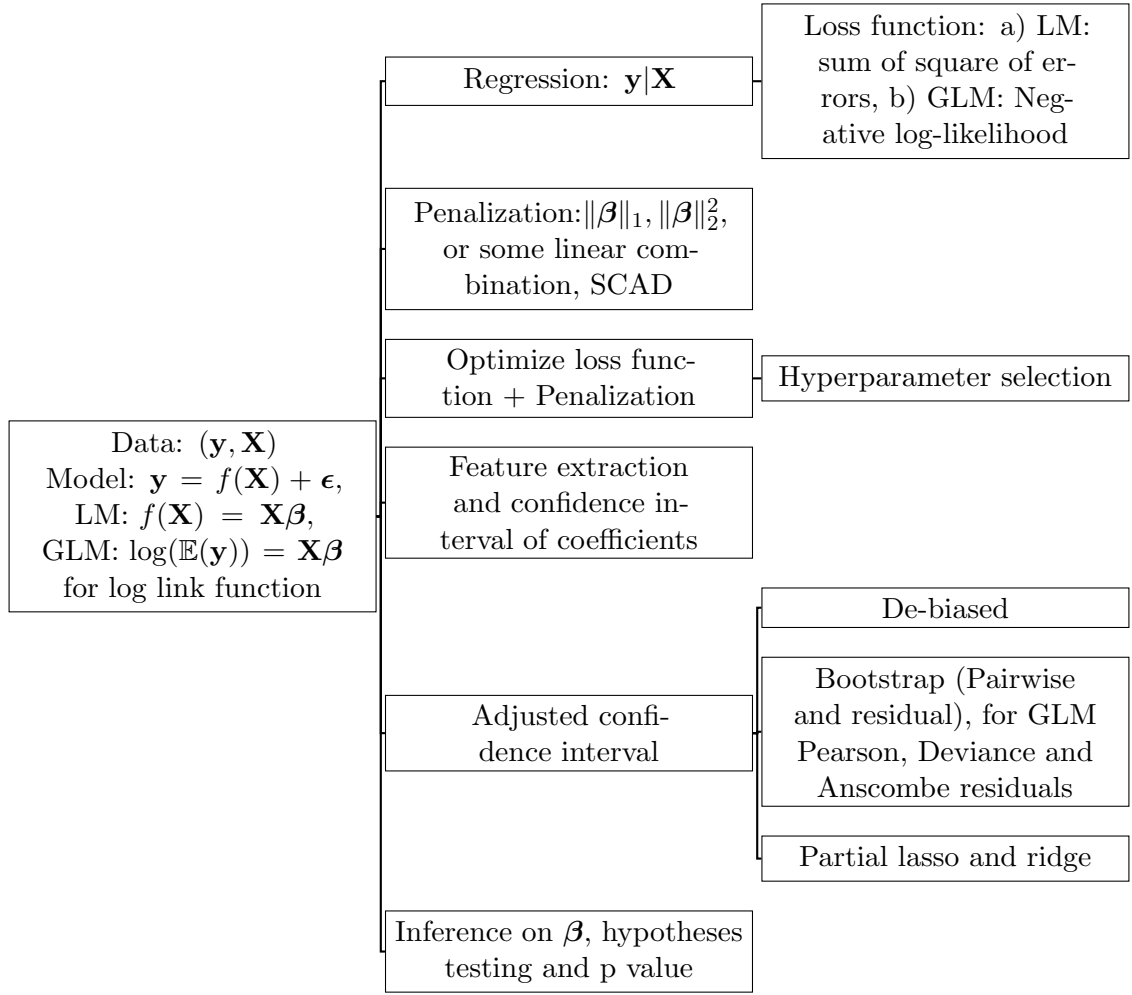
\begin{figure}[h!]
  \centering
\begin{tikzpicture}[scale=0.8, grow'=right, level distance=2.1in, sibling distance=.2in]
\tikzset{
  edge from parent/.style={thick, draw, edge from parent fork right},
  every tree node/.style={draw, minimum width=.9in, text width=2in, align=center}
}

\Tree 
[.{\textbf{Input Data:} $\left(\mathbf{y},\mathbf{X}\right)$ \\ 
   \textbf{Model:} $\mathbf{y}=f(\mathbf{X})+\epsilonvec$ \\ 
   LM: $f(\mathbf{X})= \mathbf{X}\betavec$ \\ 
   GLM: $\log(\mathbb{E}(\mathbf{y}))= \mathbf{X}\betavec$ (log link)}
    [.{\textbf{Step 1: Regression} \\ $\mathbf{y}|\mathbf{X}$} 
        [.{\textbf{Loss Function:} \\ a) LM: Sum of squared errors \\ b) GLM: Negative log-likelihood} ]
    ]
    [.{\textbf{Step 2: Penalization }\\ $\Vert\betavec\Vert_{1}$, $\Vert\betavec\Vert_{2}^{2}$, SCAD or a linear combination} ]
    [.{\textbf{Step 3: Optimization} \\ Minimize (Loss + Penalization)} 
        [.{\textbf{Hyperparameter Selection}} ]
    ]
    [.{\textbf{Step 4: Feature Extraction} \\  Confidence intervals account for feature selection bias.} ]
    [.{\textbf{Step 5: Adjusted Confidence Interval:} Methods like de-biasing and bootstrapping are applied to ensure valid inference after selection.} 
        [.{\textbf{\textbf{Step 5.1: De-biased Estimators}}} ]
        [.{\textbf{\textbf{Step 5.2: Bootstrap Methods}} \\ Pairwise and residual \\ For GLM: Pearson, Deviance, Anscombe residuals} ]
        [.{\textbf{\textbf{Step 5.3: Partial Lasso and Ridge}}} ]
    ]
   [.{\textbf{Output: \\ Inference on $\betavec$} \\ Hypothesis Testing \\ p-values, Output p-values and hypothesis tests are adjusted for the selection procedure, providing valid statistical inference.} ]
]
\end{tikzpicture}
\caption{Schematic diagram of feature extraction and selective inference for LM and GLM.}
\label{fig:feature_extraction_tree}
\footnotesize \textbf{Note:} After model fitting and penalization, selective inference starts by extracting features and constructing confidence intervals for the estimated coefficients $\betavec$. Selective inference ensures that p-values and hypothesis tests for the features are valid after feature selection. Without these adjustments, traditional p-values would not account for the selection process, leading to misleading results.
\end{figure}

We first describe the feature extraction process and re-estimation of the confidence interval in the diagram \ref{fig:feature_extraction_tree}. The detailed methodology is described below.

\begin{figure}[h!]
\centering
\tikzstyle{startstop} = [rectangle, rounded corners, 
minimum width=3cm, 
minimum height=1cm,
text centered, 
draw=black, 
fill=red!30]

\tikzstyle{io} = [trapezium, 
trapezium stretches=true, 
trapezium left angle=70, 
trapezium right angle=110, 
minimum width=3cm, 
minimum height=1cm, text centered, 
draw=black, fill=blue!30]

\tikzstyle{process} = [rectangle, 
minimum width=3cm, 
minimum height=1cm, 
text centered, 
text width=12cm, 
draw=black, 
fill=orange!30]

\tikzstyle{decision} = [diamond, 
minimum width=1cm, 
minimum height=1cm, 
text centered, 
draw=black, 
fill=green!10]
\tikzstyle{arrow} = [thick,->,>=stealth]
\begin{tikzpicture}[node distance=1.9cm]
\node (start) [startstop] {Data $\left(\mathbf{y},\mathbf{X}\right)$};
\node (in1) [io, below of=start] { $\argmin_{\bbeta}\{\mathcal{L}(\mathbf{y},\mathbf{X}\betavec)+P(\bbeta,\lambda)\}$};
\node (pro1) [process, below of=in1] {$\hat{\bbeta}$ is biased! Calculate penalized covariance of $\hat{\bbeta}$ };
\node (dec1) [decision, below of=pro1, yshift=-0.5cm] {De-biasing};

\node (pro2a) [process, below of=dec1, yshift=-0.59cm] {$\text{LM CI}=(\hat{\beta}_{j}^{\text{debias}}-1.96\sqrt{V_{j}},
    \hat{\beta}_{j}^{\text{debias}}+1.96\sqrt{V_{j}}).$\\
    $ \text{GLM CI}=(\hat{b_{j}}-1.96\sqrt{(\hat{\Theta}\Sigma\hat{\Theta}^{T})_{jj}/n} ,\hat{b_{j}}+1.96\sqrt{(\hat{\Theta}\Sigma\hat{\Theta}^{T})_{jj}/n}).$
};

\node (out1) [io, below of=pro2a] {Inference over $\bbeta$, hypotheses testing and p value};

\draw [arrow] (start) -- (in1);
\draw [arrow] (in1) -- (pro1);
\draw [arrow] (pro1) -- (dec1);
\draw [arrow] (dec1) -- node[anchor=east] {} (pro2a);
\draw [arrow] (pro2a) -- (out1);
\end{tikzpicture}
\caption{Schematic diagram of de-biased estimates for LM and GLM}
\label{debiased_lasso}
\footnotesize \textbf{Note:} 
\begin{itemize}
    \item The process starts with input data \((\mathbf{y}, \mathbf{X})\). Penalized regression methods (such as Lasso or Ridge) are used to estimate coefficients by minimizing a loss function combined with a penalty term. However, this produces biased estimates, denoted as \(\hat{\bbeta}\).
\item To address the bias introduced by penalization, a de-biasing step is performed. This step adjusts the estimates and ensures that they are suitable for inference.
\item Selective inference is incorporated by adjusting the confidence intervals to account for the selection process. This ensures valid statistical conclusions even when features are selected during the analysis.
\item The final output includes valid confidence intervals, hypothesis tests, and p-values for the coefficients, incorporating adjustments for the feature selection bias.

\end{itemize}
\end{figure}
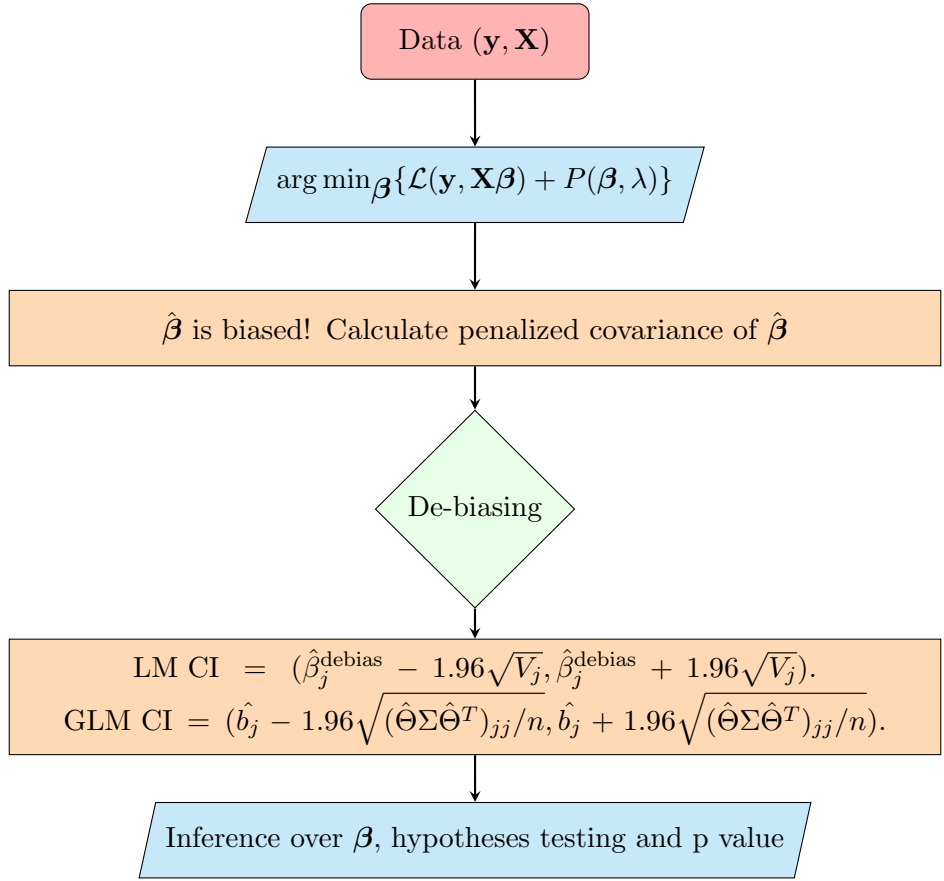

\subsection{De-biased estimator for the linear model}
The lasso estimate often has non-negligible estimation bias for large
coefficients. To have valid inference on the coefficients, one approach is to
correct the bias of the lasso estimate. Recall that the lasso estimator (see \cite{tibshirani1996regression}) is
\begin{equation}
\begin{aligned}
  \widehat{\bbeta}=\hat{\bbeta}_{n}(\Yvec,\Xvec,\lambda)
  =\argmin_{\bbeta}\frac{1}{2n}\|\Yvec-\Xvec\bbeta\|_{2}^{2}
  +\lambda\|\bbeta\|_{1}.
\end{aligned}
\end{equation}
Let \(\hat{\boldsymbol{\Sigma}}_{p\times p}=\frac{1}{n}\Xvec'\Xvec\). Note that for
\(j=1,\ldots,p\) and for columnwise vector $\uvec\in\mathbb{R}^{p}$,
\begin{equation}\label{eq:debiasing-lasso}
  \begin{aligned}
    (\hat{\beta}_{j}-\beta_{j})+
    \uvec^{T}\frac1n\Xvec'(\Yvec-\Xvec\hat{\bbeta})
    &=\uvec^{T}\frac1n\Xvec'(\Yvec-\Xvec\bbeta
      +\Xvec\bbeta-\Xvec\hat{\bbeta})
      -(\beta_{j}-\hat{\beta}_{j}) \\
    &=\uvec^{T}\frac1n\Xvec'\epsilonvec
      +(\hat{\boldsymbol{\Sigma}}\uvec-\mathbf{e}_{j})^{T}
      (\bbeta-\hat{\bbeta}),
  \end{aligned}
\end{equation}
where $\mathbf{e}_{j}$ is the vector whose $j$th coordinate is equal to one and
the rest equal to zero. The first term of the rightmost-hand side of
\eqref{eq:debiasing-lasso} has variance
$n^{-1}\sigma_{\epsilon}^{2}\uvec^{T}\hat{\boldsymbol{\Sigma}}\uvec$ conditional
on the design matrix \(\Xvec\), and the second term is bounded by
$\|\hat{\boldsymbol{\Sigma}}\uvec-\mathbf{e}_{j}\|_{\infty}\|
\bbeta-\hat{\bbeta}\|_{1}$, where $\|\mathbf{x}\|_\infty = \max_{1 \leq i \leq n} |x_i|$. Intuitively, one needs to find a \(\uvec\) such that
both the variance and
\(\|\hat{\boldsymbol{\Sigma}}\uvec-\mathbf{e}_{j}\|_{\infty}\) are small. One
way to find \(\uvec\) for each of the columns \(j=1,\ldots,p\) is through the following convex programming for some $\mu > 0$,
\begin{equation}
  \begin{split}
    \argmin_{\uvec}&\enskip\uvec^{T}\hat{\boldsymbol{\Sigma}}\uvec\\
    \text{subject to}&\enskip\|\hat{\boldsymbol{\Sigma}}\uvec
      -\mathbf{e}_{j}\|_{\infty}\leq\mu.
  \end{split}
\end{equation}
Note that we are going to solve ``p'' many optimization problems where $\mathbf{e}_{j}$ is different and defined earlier as  coordinate vector. After obtaining the optimal $\hat{\uvec}_{j}$ for \(j=1,\ldots,p\), we put them
column-wise to construct the matrix \(\hat{\boldsymbol{M}}\). The de-biased
lasso estimator is given by
\begin{equation}
  \begin{aligned}
    \hat{\bbeta}_{n}^{\text{debias}}
    =\hat{\bbeta}_{n}(\Yvec,\Xvec,\lambda)+\frac{1}{n}\hat{\boldsymbol{M}}
    \Xvec^{T}(Y-\Xvec^{T}\hat{\bbeta}_{n}(\Yvec,\Xvec,\lambda)).
  \end{aligned}
\end{equation}
It can be shown that asymptotically \(\hat{\beta}_{j}^{\text{debias}}\) follows
a normal distribution centered around \(\beta_{j}^{*}\), the true value of the ``j'' th coefficient. The \(95\%\)
confidence interval for a component \(\beta_{j}\) can  thus be constructed as
\begin{equation}
  \begin{aligned}
    \text{CI}=(\hat{\beta}_{j}^{\text{debias}}-1.96\sqrt{V_{j}},
    \hat{\beta}_{j}^{\text{debias}}+1.96\sqrt{V_{j}}),
  \end{aligned}
\end{equation}
where \(V_{j}=n^{-1}\hat{\sigma}_{\epsilon}^{2}\hat{\uvec}_{j}^{T}
\hat{\Sigma}\hat{\uvec}_{j}\). See more details on error bounds and detailed simulation studies in \cite{van2014asymptotically}, \cite{javanmard2014confidence}, and \cite{cai2023statistical}. The bias in confidence interval estimation tends to 0 when the sample size n is in the order of $(\log(p))^{2}$ where p is the dimension of the data whereas for estimation and prediction purposes we need the sample size in the order of $\log(p)$. A flowchart is given in the diagram $\ref{debiased_lasso}$. 

\subsection{De-biased estimators for the GLM}

More generally, we consider the elastic net penalized in Generalized Linear Model (GLM), which combines the properties of Lasso and Ridge regression by incorporating both \( \ell_1 \) and \( \ell_2 \) penalties. It is important to note that we consider the minimization of the negative of log-likelihood for the GLM. The elastic net estimator is defined as:
\begin{equation}
  \begin{aligned}
    \hat{\bbeta}=\argmin_{\bbeta}
    \ell(\bbeta)+\lambda_{1}\Vert\bbeta\Vert_{1}
    +\lambda_{2}\Vert\bbeta\Vert_{2}^{2},
  \end{aligned}
\end{equation}
where
\[
\ell(\bbeta)=-\frac{1}{n}\sum_{i=1}^{n}w_{i}\bigl[y_{i}h(\bx_{i}^{T}\bbeta)
  -b\bigl(h(\bx_{i}^{T}\bbeta)\bigr)\bigr]
\]
is the negative log-likelihood scaled by the sample size \(n\). The de-biased estimator is defined as 
\[
\label{eqn:debiased_eq}
\hat{\mathbf{b}} = \hat{\bbeta} - \widehat{\Theta} \nabla \ell(\hat{\bbeta}),
\]
where \( \widehat{\Theta} = (\nabla^2 \ell(\hat{\bbeta}))^{-1} \) is the inverse of the Hessian matrix of the loss function \( \ell(\cdot) \), provided it exists. This formulation assumes \( p \leq n \), ensuring that the Hessian is invertible. In the high-dimensional scenario (\emph{i.e.},
\(p>n\)), \(\widehat{\Theta}\) is often some (sparse) approximation to
\(E[\nabla^{2}\ell(\bbeta)]\). Under regularity conditions, $\hat{\mathbf{b}}$
can be shown to follow an asymptotic normal distribution. Denote $\bbeta^{*}$
the true coefficient vector. Observe that
\begin{equation}
  \begin{aligned}
  \label{eqn:taylor_series}
    &\nabla\ell(\bbeta^{*})\\
    &=\nabla\ell(\hat{\bbeta})-\nabla^{2}\ell(\bbeta^{*})
      (\hat{\bbeta}-\bbeta^{*})-r(\Vert(\hat{\bbeta})
      (\hat{\bbeta}-\bbeta^{*})\Vert_{2})\\
    &=\nabla^{2}\ell(\bbeta^{*})(\bbeta^{*}-\hat{\bbeta}-(\nabla^{2}\ell(\hat{\bbeta}))^{-1}l^{'}(\hat{\bbeta}))-r(||(\hat{\bbeta})(\hat{\bbeta}-\bbeta^{*})||_{2})\\
                        &=\nabla^{2}\ell(\bbeta^{*})(\bbeta^{*}-\hat{\bbeta}+\hat{\Theta}l^{'}(\hat{\bbeta}))-\nabla^{2}\ell(\bbeta^{*})((\nabla^{2}\ell(\bbeta^{*}))^{-1}+\hat{\Theta})(\nabla\ell(\hat{\bbeta}))-r(||(\hat{\bbeta})(\hat{\bbeta}-\bbeta^{*})||_{2})\\
                        &=: \nabla^{2}\ell(\bbeta^{*})(\bbeta^{*}-\hat{\bbeta}+\hat{\Theta}l^{'}(\hat{\bbeta}))+R_{n}.
  \end{aligned}
\end{equation}
$r(.)$ is the residual function in the Taylor-series expansion and \[R_{n}:=-\nabla^{2}\ell(\bbeta^{*})((\nabla^{2}\ell(\bbeta^{*}))^{-1}+\hat{\Theta})(\nabla\ell(\hat{\bbeta}))-r(||(\hat{\bbeta})(\hat{\bbeta}-\bbeta^{*})||_{2}).\]
Now under certain regularity conditions and using equation \ref{eqn:debiased_eq} and \ref{eqn:taylor_series}, we can achieve normal distribution as given below.
\begin{equation}
    \begin{aligned}
       \sqrt{n}(\hat{\bf{b}}-\bbeta^{*})\approx \hat{\Theta} \sqrt{n}(R_{n}-\ell^{'}(\bbeta^{*}))\xrightarrow{d}N(0,\hat{\Theta}\Sigma\hat{\Theta}^{T}),
    \end{aligned}
\end{equation}
where $\Sigma=\textrm{asymptotic variance of } \sqrt(n)\textrm{Var}(\ell^{'}(\bbeta^{*}))$. Hence the confidence interval will be found as
\begin{equation}
    \begin{aligned}
        \text{CI}=(\hat{b_{j}}-1.96\sqrt{(\hat{\Theta}\Sigma\hat{\Theta}^{T})_{jj}/n} ,\hat{b_{j}}+1.96\sqrt{(\hat{\Theta}\Sigma\hat{\Theta}^{T})_{jj}/n}).\\
        \end{aligned}
\end{equation}
For details please see \cite{zhang2014confidence}.

\subsection{Bootstrap estimation in linear models}
 In this section, we will elaborate on some common practices for the bootstrap confidence interval. Let
$\Yvec=(y_{1},y_{2},\cdots,y_{n})^{T}$ and
$\Xvec=(\bx_{1}^{T},\bx_{2}^{T},\cdots,\bx_{n}^{T})^{T}$. Here we consider a simple linear model $y_{i}=\bx_{i}^{T}\bbeta+\epsilon_{i}$, where $\text{E}(\epsilon_{i})=0$ and $\text{Var}(\epsilon_{i})=\sigma_{\epsilon}^{2}$. Here the estimation of $\bbeta$ is
\begin{equation}
    \begin{aligned}
        \hat{\bbeta_{\rho}}=\argmin_{ \bvec \in \mathbb{R}^{p}} \Sigma_{i=1}^{n}\rho(y_{i}-\bx_{i}^{T}\bvec)
    \end{aligned}
\end{equation}
for different choice of convex function $\rho$ such as $\rho(x)=x^2$, $\rho(x)=|x|$, Huber Loss $\rho(x)=\frac{x^2}{2}1_{|x|<k}+(k|x|-\frac{k^2}{2})1_{|x|\geq k}$ e.t.c. Now there are two common bootstrap methods below.\\
\textbf{Residual Bootstrap: }
We first obtain the errors after fitting the model and get the errors $\hat{\epsilon}_{i}=y_{i}-\bx_{i}^{T}\hat{\bbeta}$.
In this approach, we resample the errors \( \epsilon^{*}_i \) from the set of residuals \( \left\{\hat{\epsilon}_1, \hat{\epsilon}_2, \ldots, \hat{\epsilon}_n \right\} \), and form the new data points as \( y^{*}_i = \mathbf{x}_i^T \hat{\bbeta_{\rho}} + \epsilon^{*}_i \). We then re-estimate the coefficients \( \hat{\bbeta_{\rho}}^{*b} \), for \( b = 1, 2, \ldots, B \), and draw inference based on these resampled estimates. In this model, we assume a linear structure for the expected value of \( y_i \), i.e., \( \text{E}(y_i) \), and the design matrix \( \Xvec \) is treated as fixed.
\\
\textbf{Paired bootstrap: }In this method we do resample $(y_{i},\bx_{i}^{T})\in \mathbb{R}^{p+1}$ given by the empirical joint distribution $\{(y_{i},\bx_{i}^{T})\}_{i=1}^{n}$. It does not assume any mean structure and as we do resample the $\bx_{i}^{T}$, the design matrix is not fixed.

\subsection{Bootstrap estimation in linear models with penalization}
Let $\{a_{n}\}$ be a sequence of positive numbers such that $a_{n}+\frac{\log(n)}{a_{n}\sqrt(n)}\rightarrow0$ as $n\rightarrow \infty$. Let $\hat{\betavec_{\lambda}}$ be an original lasso estimate and $\hat{\betavec}^{modified}$ is another modified estimate based on the lasso estimates, defined as $\hat{\beta_{j}^{modified}}=\hat{\beta_{\lambda j}}I\{|\hat{\beta_{\lambda j}}|<a_{n}\}$ for all coordinates $j=1,2,\cdots,p$. For more details please refer \cite{chatterjee2010asymptotic}.
Now the steps are described in Algorithm \ref{alg:lasso_bootstrap_lm}.
\begin{algorithm}[t]
\caption{Lasso with Bootstrap for Confidence Intervals}
\label{alg:lasso_bootstrap_lm}
\begin{algorithmic}[1] 
  \State  Select the optimal $\hat{\lambda}$ via cross-validation and obtain the corresponding estimate $\hat{\boldsymbol{\beta}}_{\hat{\lambda}}$.
   \State   Calculate the modified estimate $\hat{\boldsymbol{\beta}}^{\text{modified}}$.
    \State  Calculate the modified residuals $\mathbf{e}^{\text{modified}}$.
    \State  Bootstrap the modified residuals $(e_{1}^{**}, e_{2}^{**}, \dots, e_{n}^{**})$ and set $y_{i}^{**} = \mathbf{x}_{i}^{T} \hat{\boldsymbol{\beta}}^{\text{modified}} + e_{i}^{**}$.
    \State  Perform lasso on the bootstrapped data to obtain the coefficients $\hat{\boldsymbol{\beta}}_{\hat{\lambda}}^{**}$.
    \State  Repeat steps 4-5 multiple times to obtain the quantiles for $\hat{\boldsymbol{\beta}}_{\hat{\lambda}}^{**}$.
   \State  Compute the confidence interval as:
    \[
    \left[\hat{\boldsymbol{\beta}}_{\hat{\lambda}} + \hat{\boldsymbol{\beta}}^{\text{modified}} - \hat{\boldsymbol{\beta}}_{\widehat{\lambda(1-\frac{\alpha}{2})}^{**}}, \hat{\boldsymbol{\beta}}_{\hat{\lambda}} + \hat{\boldsymbol{\beta}}^{\text{modified}} - \hat{\boldsymbol{\beta}}_{\widehat{\lambda(\frac{\alpha}{2})}^{**}}\right], \text{or },
    \]
    \[\left[2\hat{\boldsymbol{\beta}}_{\hat{\lambda}} - \hat{\boldsymbol{\beta}}_{\widehat{\lambda(1-\frac{\alpha}{2})}^{**}}, 2\hat{\boldsymbol{\beta}}_{\hat{\lambda}} - \hat{\boldsymbol{\beta}}_{\widehat{\lambda(\frac{\alpha}{2})}^{**}}\right]. 
    \]
\end{algorithmic}
\end{algorithm}
\subsection{Bootstrap estimation in GLM with penalization}
The Algorithm as mentioned earlier \ref{alg:lasso_bootstrap_lm} is for the linear model. If we want to extend the methodology for GLM, we need to bring a counterpart of errors such as Pearson residuals. 
\begin{algorithm}[t]
\caption{Lasso with residual Bootstrap for Confidence Intervals for GLM}
\label{alg:lasso_bootstrap_glm}
\begin{algorithmic}[1] 
\State  Select optimum $\hat{\lambda}$ via cross validation and the corresponding $\hat{\betavec}_{\hat{\lambda}}$.
    \State Calculate the modified estimate.
    \State Calculate Pearson residuals $R_{i}^{P}=\frac{Y_{i}-\widehat{f}_{n}(\bX_{i})}
    {\sqrt{v_{i}}},
    i\in\{1,2,\cdots,n\}.$
    For Tweedie, $\sqrt{v_{i}}=\widehat{\phi}^{1/2}[\widehat{f}_{n}(\bX_{i})]^{p/2}$, for Poisson $\sqrt{v_{i}}=\sqrt{\widehat{f}_{n}(\bX_{i})}$, for Negative-Binomial $v_i = \widehat{f}_n(\bX_{i}) + \frac{\widehat{f}_n(\bX_{i})^2}{\widehat{\theta}}
$, where $\theta$ is the dispersion parameter e.t.c.
    \State Bootstrap the Pearson residuals $(e_{1}^{**},e_{2}^{**},\cdots,e_{n}^{**})$ and set $y_{i}^{**}=\sqrt{v_{i}}e_{i}^{**}+\hat{\mu_{i}}$ where $\hat{\mu_{i}}=\widehat{f}_{n}(\bX_{i})$. Based on that again do lasso for another time to get the coefficient $\hat{\betavec}_{\hat{\lambda}}^{**}$. Now if you do the above process several times you can get the quantiles for $\hat{\betavec}_{\hat{\lambda}}^{**}$.
    \State Find $[\hat{\betavec}_{\lambda}+\hat{\betavec}^{modified}-\hat{\betavec}_{\widehat{\lambda (1-\frac{\alpha}{2})}^{**}};\hat{\betavec}_{\lambda}+\hat{\betavec}^{modified}-\hat{\betavec}_{\hat{\lambda}(\frac{\alpha}{2})}^{**}]$ or $[2\hat{\betavec}_{\lambda}-\hat{\betavec}_{\widehat{\lambda (1-\frac{\alpha}{2})}^{**}};2\hat{\betavec}_{\lambda}-\hat{\betavec}_{\widehat{\lambda}(\frac{\alpha}{2})}^{**}]$ where $\hat{\beta_{j}^{modified}}=\hat{\beta_{\lambda j}}I\{|\hat{\beta_{\lambda j}}|<a_{n}\}$ for each of the coordinates of the vector $\hat{\betavec}^{modified}$.
\end{algorithmic}
\end{algorithm}
The Algorithm \ref{alg:lasso_bootstrap_glm} is based on residual bootstrap. One can do a pair bootstrap variant for the above Algorithm. For Tweedie regression, we can apply this method.  Recall that the power parameter \(p\in(1,2)\). Letting $\hat{\phi}$ to be the dispersion parameter and $\widehat{f}_{n}(\bX_{i})=\hat{\mu}_{i}$ the Pearson residuals can be written as 
  \begin{equation*}
    R_{i}^{P}=\frac{Y_{i}-\widehat{f}_{n}(\bX_{i})}
    {\widehat{\phi}^{1/2}[\widehat{f}_{n}(\bX_{i})]^{p/2}},
    i\in\{1,2,\cdots,n\}.
  \end{equation*} 
Deviance residuals can be written as
  \begin{equation*}
    R_{i}^{D}=\text{sgn}(Y_{i}-\widehat{f}_{n}(\bX_{i}))
    \sqrt{2\left(\frac{Y_{i}[\widehat{f}_{n}(\bX_{i})]^{1-p}}{p-1}
      -\frac{Y_{i}^{2-p}}{(p-1)(2-p)}
      +\frac{[\widehat{f}_{n}(\bX_{i})]^{2-p}}{2-p}\right)},
    i\in\{1,2,\cdots,n\}.
    \end{equation*}
Another important residual commonly known as Anscombe residual is
  \begin{equation*}
    R_{i}^{A}=\frac{\frac{3}{3-p}(Y_{i}^{1-p/3}-[\widehat{f}_{n}(\bX_{i})]^{1-p/3})}
    {[\widehat{f}_{n}(\bX_{i})]^{p/6}},i\in\{1,2,\cdots,n\}.
  \end{equation*} We explained in Algorithm \ref{alg:lasso_bootstrap_glm} for Pearson residuals. One can also extend the idea for Deviance and Anscombe residuals (see \cite{pierce1986residuals}) in practice when we have a significant departure from symmetry.
\subsection{Paired bootstrap lasso with partial ridge estimation } \label{plr_lm}
In this section, we shall discuss a variant of the bootstrap lasso which includes a partial ridge estimation to avoid over-shrinkage of the large estimates. Please see \cite{liu2020bootstrap} for more details on lasso with partial ridge. We may use the idea for the GLM.
\begin{algorithm}[t]
\caption{Paired Bootstrap with Lasso and Ridge Regression}
\label{alg:paired_bootstrap}
\begin{algorithmic}[1]
\State  Take a paired bootstrap sample.
\State  Find the Lasso estimate of the coefficients on the bootstrap sample.
\State  Identify the indices of the zero coefficients from the Lasso estimate.
\State  Perform ridge regression on the bootstrap paired data with $\ell_{2}$ penalty, applied only to the coefficients identified as zero in the previous step.
\State If you perform steps 1 to 4, you will obtain one estimate of $\hat{\bbeta}_{b}$. Repeat steps 1 to 4 a total of $B$ times to obtain quantiles of the estimates from $\hat{\bbeta}_{b};b=1,\cdots,B$ for each of the coordinates.
\end{algorithmic}
\end{algorithm}
For the simulation studies, we have seen that for Poisson regression and Logistic regression, the paired bootstrap gives more promising results than the residual bootstrap. A paired bootstrap lasso with a partial ridge gives better results than only a paired bootstrap strategy.
\subsection{Penalization over Tweedie Regression}\label{p_tweedie}

In the context of de-biasing the LASSO estimator of $\bm{\theta}^*$, the true value of the parameter $\thetavec$, an alternative approach has been proposed by \cite{van2014asymptotically}, which is the declassified estimator of $\bm{\theta}^*$. This method is more general than the de-biased estimator of \cite{javanmard2014confidence} because it extends readily to $l_1$ regularized M-estimators. Let $\rho(y_i, b)$ denote a convex loss function in $b$, and let $\rho'$ and $\rho''$ denote its first and second derivatives with respect to $b$, respectively, as
$$\rho'(y_i, b)=\frac{\partial}{\partial b}\rho(y_i, b), \rho''(y_i, b)=\frac{\partial^2}{\partial b^2}\rho(y_i, b).$$
We denote the loss function $\ell(\bm{\theta}) = \frac{1}{n} \sum_{i=1}^n\rho(y_i, \beta_0 +  \bm{\beta}^T \mathbf{x}_i)$. The de-sparsified estimator can be expressed as follows:
\begin{equation}
    \hat{\bm{\theta}}^d = \hat{\bm{\theta}}_{l} - \hat{\bm{\Theta}} \nabla \ell (\hat{\bm{\theta}}),
\end{equation}
where $\hat{\bm{\theta}}_{l}$ is the local lasso estimator with the form
$$\hat{\bm{\theta}}_l = \argmin_{(\beta_0, \bm{\beta})} \frac{1}{n} \sum_{i=1}^n \bigg( \frac{y_i \exp((1-\rho)(\beta_0 + \bm{\beta}^T \mathbf{x}_i))}{1-\rho} + \frac{\exp((2-\rho)(\beta_0 + \bm{\beta}^T \mathbf{x}_i))}{2-\rho} \bigg) + \lambda_k ||\bm{\beta}||_1$$
and $\hat{\bm{\Theta}}$ is a regularized inverse of the Hessian matrix of second-order derivatives of loss function given the corresponding estimator which is the approximate inverse of $\mathbf{\Sigma}$ \citep{van2014asymptotically}. There are two ways to calculate the $\hat{\bm{\Theta}}$.
\begin{itemize}
    \item \cite{xia2020revisit} suggest that when $p < n$, it is preferred to use the refined de-biased lasso, which directly inverts the Hessian matrix and provides improved confidence interval coverage probabilities for a wide range of $p$. 
    \item Column-by-column estimation methods are the exact regularized inverse of the Hessian matrix.
    \begin{enumerate}
        \item Nodewise Lasso estimator for the estimated Hessian matrix \citep{Meinshausen2006}
        \item Constrained $\ell_1$-Minimization for Inverse Matrix Estimation (CLIME) \citep{tony2011}
        \item Sparse Column-wise Inverse Operator (SICO) \citep{Liu2012HD}
        \item Scaled-Lasso methods \citep{sun2012zhang}
    \end{enumerate}
\end{itemize}

For the nodewise method, to estimate $\hat{\bm{\Theta}}$, we consider the lasso type optimization:
\begin{equation}
    \hat{\bm{\gamma}}_j = \argmin_{\bm{\gamma}} ( \hat{\bm{\Sigma}}_{j,j} - 2\hat{\bm{\Sigma}}_{j,-j}\bm{\gamma} + \bm{\gamma}^T \hat{\bm{\Sigma}}_{-j,-j} \bm{\gamma} + 2\lambda_j ||\bm{\gamma}||_1), 
\end{equation}
where $\hat{\bm{\Sigma}}_{j,-j}$ is the $j$th row of $\hat{\bm{\Sigma}}$ without the $j$th element, and $\hat{\bm{\Sigma}}_{-j, -j}$ is the sub-matrix of  is the sub-matrix of $\hat{\bm{\Sigma}}$ without the $j$th row and $j$th column for $j = 1, 2, ..., d$. Based on (21), we have $\hat{\tau}_j^2 = \hat{\bm{\Sigma}}_{j,j} - \hat{\bm{\Sigma}}_{j,-j} \hat{\bm{\gamma}}_j$, $\hat{\bm{T}}=\diag(\hat{\tau}_1, \hat{\tau}_2, \dots, \hat{\tau}_d)$ and 
\begin{equation}
    \hat{\bm{C}} = 
    \begin{pmatrix}
        1 & -\hat{\gamma}_{1,2} & \cdots & -\hat{\gamma}_{1,d} \\
        -\hat{\gamma}_{2,1} & 1 & \cdots & -\hat{\gamma}_{2,d} \\
        \vdots  & \vdots  & \ddots & \vdots  \\
        -\hat{\gamma}_{d,1} & -\hat{\gamma}_{d,2} & \cdots & 1 
    \end{pmatrix}.
\end{equation}
Then we can define $\hat{\bm{\Theta}}$ as
\begin{equation}
    \hat{\bm{\Theta}} = \hat{\bm{T}}^{-1} \hat{\bm{C}}.
\end{equation}

For Tweedie GLM, the sample covariance matrix based on the data can be written as $\hat{\bm{\Sigma}}_{\hat{\bm{\beta}}} = \mathbf{X}_{\hat{\bm{\beta}}}^T \mathbf{X}_{\hat{\bm{\beta}}}/n$, where the weighted design matrix $\mathbf{X}_{\hat{\bm{\beta}}} = \bm{W}_{\hat{\bm{\beta}}} \mathbf{X}$ and $\bm{W}_{\hat{\bm{\beta}}}$ is the diagonal matrix as
$$(\bm{W}_{\hat{\bm{\beta}}})_{i,i} = \rho''(y_i, \beta_0 + \bm{\beta}^T \mathbf{x}_i).$$
The advantage is we can simply use the nodewise lasso based on the weighted design matrix and it becomes easier to implement by R package \textbf{glmnet}. In particular, we can directly apply lasso linear regression
\begin{equation}
    \hat{\bm{\gamma}}_j = \argmin_{\bm{\gamma}} \frac{1}{2n} ||\mathbf{X}_{\hat{\bm{\beta}},j} - \mathbf{X}_{\hat{\bm{\beta}},-j}\bm{\gamma}||_2^2 + \lambda_j ||\bm{\gamma}||_1,
\end{equation}
where $\bm{\gamma} \in \mathbb{R}^p$, $\mathbf{X}_{\hat{\bm{\beta}},j}$ is $j$th column of the weighted design and $\mathbf{X}_{\hat{\bm{\beta}},-j}$ is the weighted design without the $j$th column. So the $j$th row of $\hat{\bm{\Theta}}$ is
\begin{equation}
    \hat{\bm{\Theta}}_{j,.} = \frac{1}{\hat{\tau}_j}
    \begin{bmatrix}
        -\hat{\gamma}_{j,1} & \cdots & -\hat{\gamma}_{j,j-1} & 1 & -\hat{\gamma}_{j,j+1} & \cdots & -\hat{\gamma}_{j,p}
 \end{bmatrix},
\end{equation}
where 
$$\hat{\tau}_j = \sqrt{\bigg( \frac{1}{n} ||\mathbf{X}_{\hat{\bm{\beta}},j} - \mathbf{X}_{\hat{\bm{\beta}},-j}\bm{\gamma}||_2^2 + \lambda_j ||\bm{\gamma}||_1 \bigg)}.$$
On the other hand, by the directly invert method, $\hat{\bm{\Theta}}$ becomes $\hat{\bm{\Sigma}}_{\hat{\bm{\beta}}}^{-1}$. Besides, the 95\% confidence interval can be found as
\begin{equation}
    \begin{aligned}
        \text{CI}=(\hat{\bm{\beta}}_j -1.96\sqrt{\phi(\hat{\bm{\Theta}}\hat{\bm{\Sigma}}\hat{\bm{\Theta}}^{T})_{jj}/n},\hat{\bm{\beta}}_j+1.96\sqrt{\phi(\hat{\bm{\Theta}}\hat{\bm{\Sigma}}\hat{\bm{\Theta}}^{T})_{jj}/n}).\\
        \end{aligned}
\end{equation}

\subsection{LightGBM for variable selection}
The Tweedie loss function is used in generalized linear models (GLMs) when modeling distributions from the Tweedie family. It is defined as:

\[
\mathcal{L}(y_i, \mu_i) = \frac{1}{p-1} \left[ y_i \mu_i^{1-p} - \frac{\mu_i^{2-p}}{2-p} \right], \quad \text{for } p \neq 1, 2,
\]

where \(y_i\) is the observed response, \(\mu_i\) is the predicted mean, linked to the linear predictor via the inverse link function, \(p\) is the Tweedie index parameter, which determines the specific distribution: \(p = 0\): Normal distribution, \(p = 1\): Poisson distribution, \(p = 2\): Gamma distribution, and \(1 < p < 2\): Compound Poisson-Gamma distribution.

The aggregate loss for a dataset with \(n\) observations is given by:

\[
\mathcal{L}(\mathbf{y}, \boldsymbol{\mu}) = \frac{1}{n} \sum_{i=1}^n \frac{1}{p-1} \left[ y_i \mu_i^{1-p} - \frac{\mu_i^{2-p}}{2-p} \right].
\]

The Tweedie loss is used in the context of LightGBM (Light Gradient-Boosting Machine, see \cite{ke2017lightgbm} for details) for regression tasks, where the target variable exhibits properties of both continuous and discrete distributions, specifically when dealing with zero-inflated data or data with a high variance. This is one of the famous tree-based machine-learning approaches for great efficiency and memory consumption. We assume the
following Tweedie model on \((Y,\bX)\):
\begin{equation*}
  f_{Y}(y|\bs{x},\phi,p)=\exp\left\{\frac{1}{\phi}\left(
      \frac{y[G(\bs{x})]^{1-p}}{1-p}-\frac{[G(\bs{x})]^{2-p}}{2-p}\right)
    +c(y,\phi,p)\right\},
\end{equation*}
where s is an appropriately chosen constant and 
\begin{equation*}
  c(y,\phi,p)=\left\{
    \begin{array}{ll}
      0, & y = 0\\
      \log\left[\frac1y\sum_{j=1}^{\infty}\frac{y^{j\alpha}}{\phi^{j(1+\alpha)}(2-p)^{j}(p-1)^{j\alpha}j!\Gamma(j\alpha)}\right], & y>0
    \end{array}
  \right.
\end{equation*}
and \(\alpha=(2-p)/(p-1)\). We are mainly interested in the
regression function \(G(\bs{x})=\E(Y|\bs{X}=\bs{x})\). The power parameter
\(p\in(1,2)\) and the dispersion parameter \(\phi>0\) are nuisance. In that case, functional gradient is \[-y\exp[(1-p)G(\bx)]+\exp[(2-p)G(\bx)]\] and the corresponding hessian is \[-y(1-p)\exp[(1-p)G(\bx)]+(2-p)\exp[(2-p)G(\bx)]\] when $s=1$. The number of times a feature is used to split the data across all trees and the improvement of model accuracy after splitting across a variable are the two criteria used for evaluating feature importance in LightGBM. Typically bootstrap methods are evaluated to quantify the uncertainty of the prediction interval. 

\subsection{Bayesian perspective for credible interval}
In a Bayesian regression framework, we can compute credible intervals for the regression coefficients using a spike-and-slab prior (see \cite{ishwaran2005spike}). This prior is particularly useful for variable selection because it can represent a belief that some coefficients are exactly zero, while others are not.
The spike-and-slab prior is a mixture of priors, typically defined as:
\[
\beta_j \sim \pi \delta_0 + (1-\pi) \text{Slab}(\beta_j \mid \theta),
\]
where \(\pi\) is the prior probability that the coefficient \(\beta_j\) is zero, \(\delta_0\) is a Dirac delta function centered at zero, representing the "spike" part, and \(\text{Slab}(\beta_j \mid \theta)\) is a continuous distribution (e.g., a normal distribution \(N(0, \sigma^2)\)) representing the "slab" part, where \(\theta\) represents the hyperparameters.
Given the prior and the likelihood of the data, the posterior distribution of each coefficient \(\beta_j\) can be derived using Bayes' theorem:
\[
p(\beta_j \mid \mathbf{y}, \mathbf{X}) \propto p(\mathbf{y} \mid \mathbf{X}, \beta_j) \cdot \left[\pi \delta_0 + (1-\pi) \text{Slab}(\beta_j \mid \theta)\right],
\]
where \(\mathbf{y}\) is the response vector and \(\mathbf{X}\) is the matrix of predictors.

The credible interval for a coefficient \(\beta_j\) can be computed from its posterior distribution. For a chosen credibility level \(1-\alpha\), the interval \([L_j, U_j]\) is defined such that:
\[
P(L_j \leq \beta_j \leq U_j \mid \mathbf{y}, \mathbf{X}) = 1 - \alpha.
\]
Typically, these intervals are constructed using the posterior quantiles:
\[
L_j = F^{-1}_{\beta_j}(0.025), \quad U_j = F^{-1}_{\beta_j}(0.975),
\]
where \(F^{-1}_{\beta_j}(\cdot)\) is the inverse cumulative distribution function of the posterior distribution of \(\beta_j\).
The spike-and-slab prior allows for automatic variable selection within the Bayesian framework. If the posterior probability of \(\beta_j = 0\) is high (greater than a threshold, say 0.5), this suggests that the corresponding variable is not significant. For those coefficients with low posterior probability of being zero, their credible intervals provide insight into the magnitude and uncertainty of their effects. \cite{carvalho2009handling} discusses the implementation of "horseshoe" prior with global and local shrinkage with a connection to penalized regression. \cite{goh2018bayesian} investigated variable selection with Gaussian and diffused-gamma prior and a connection to $l_{0}$ norm penalization. 

\section{Methodology} \label{plr_glm}
In claim data, the dimension of the data is moderately large with respect to the sample size. In \cite{dawn2025some}, one may encounter a high dimension low sample size data whereas the claim data is big for the large customer database with moderate size available variables corresponding to the customers. Here we describe our proposed methodology for the confidence interval estimation of coefficients in GLM. \cite{liu2020bootstrap} described their methodology for bootstrap lasso and then a follow-up ridge regression to obtain the confidence interval of the coefficients. We will extend their methodology for the GLM setup. We will be using $\log$ link unless otherwise mentioned. The steps are we first perform lasso regression to obtain the most important variables. Now the nonzero coefficients are possible to get over-penalized by lasso penalty. Hence we keep only ridge penalty over the coefficients which are zero with $l_{1}$ penalty. After performing the ridge regression, we calculate the coefficients. If we perform these steps for $B$ many bootstrap samples, we can construct a confidence interval for each coefficient based on their 25\% and 95\% quantiles. 
\begin{algorithm}
\caption{PLR Algorithm for GLM}
\label{alg:plr_algo}
\begin{algorithmic}[1]
    \REQUIRE Data:$(y_{i},\Xvec_{i})_{i=1}^{n}$, $\Xvec_{i} \in \mathbb{R}^{p}$, $\mathcal{C}:=\{1, 2, \cdots,p\}$
    \ENSURE 95\% coverage of the coefficients
    \For{$B$ iterations}
        \State Choose a paired bootstrap sample $\mathcal{S}$ from $\{1, 2, \ldots, n\}$
        \State Perform a lasso regression on $(y_{i}, \Xvec_{i})_{i \in \mathcal{S}}$ in the proper GLM family 
        \State $\mathcal{C} = \mathcal{C}_{1} \cup \mathcal{C}_{2}$, where $\mathcal{C}_{1} := \{i \in \mathcal{P} \mid \hat{\beta}_{i} \neq 0\}$ and $\mathcal{C}_{2} := \{i \in \mathcal{P} \mid \hat{\beta}_{i} = 0\}$
        \State Perform a ridge regression on $(y_{i}, \Xvec_{i})_{i \in \mathcal{S}}$ with the $l_{2}$ penalty only over the coefficients in $\mathcal{C}_{2}$ in the proper GLM family
    \EndFor
    \RETURN After all bootstrap simulations, calculate the 95\% coverage of the coefficients.
\end{algorithmic}
\end{algorithm}

\section{Simulation studies}\label{simulation}
\subsection{Synthetic data}
\label{Synthetic data}
\begin{figure}[h]
    \centering
    \begin{subfigure}[b]{0.45\textwidth}
        \centering
        \includegraphics[width=\textwidth]{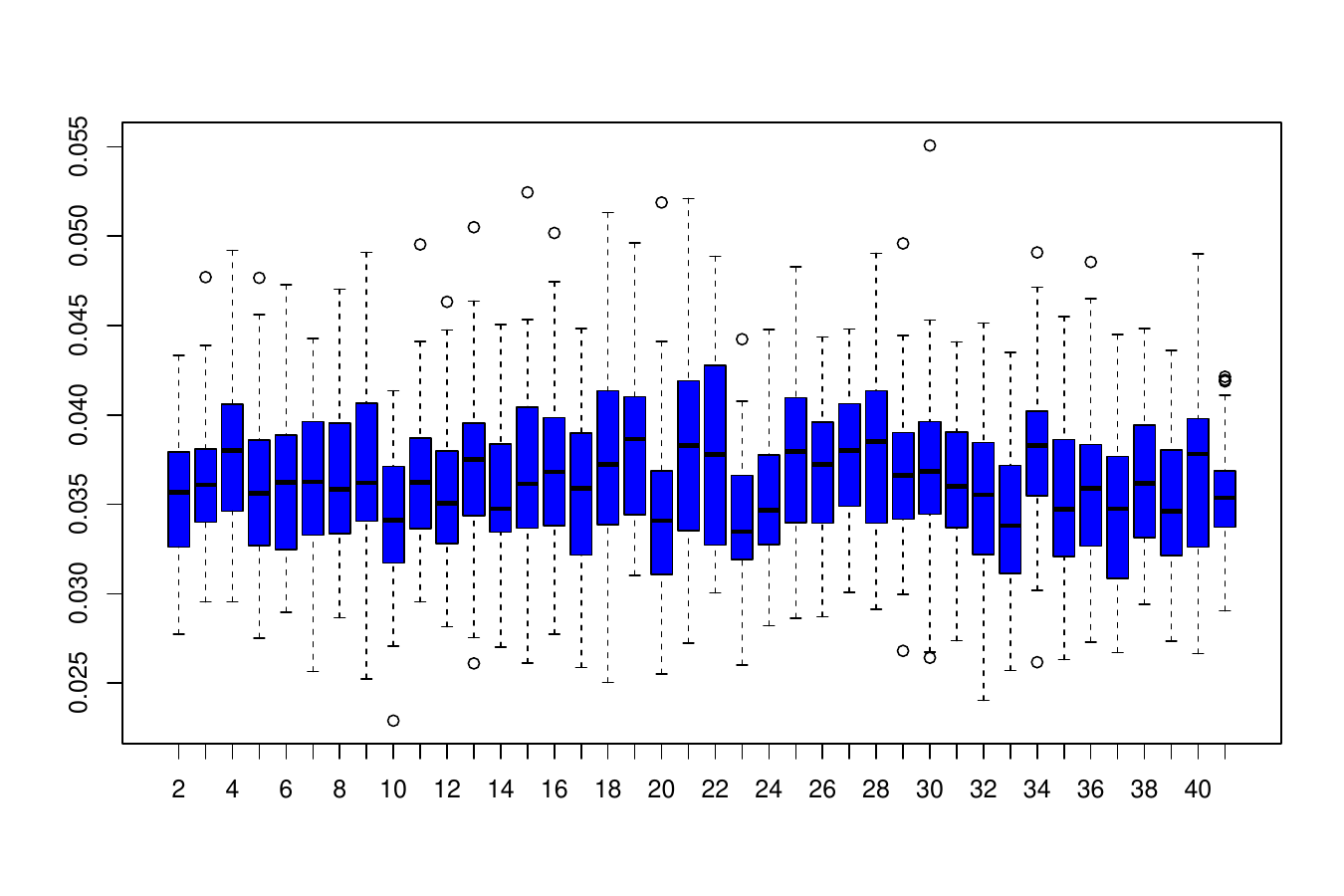}
        \caption{Negative binomial CI width box plot}
        \label{fig:nd_boxplot}
    \end{subfigure}
    \hfill
    \begin{subfigure}[b]{0.45\textwidth}
        \centering
        \includegraphics[width=\textwidth]{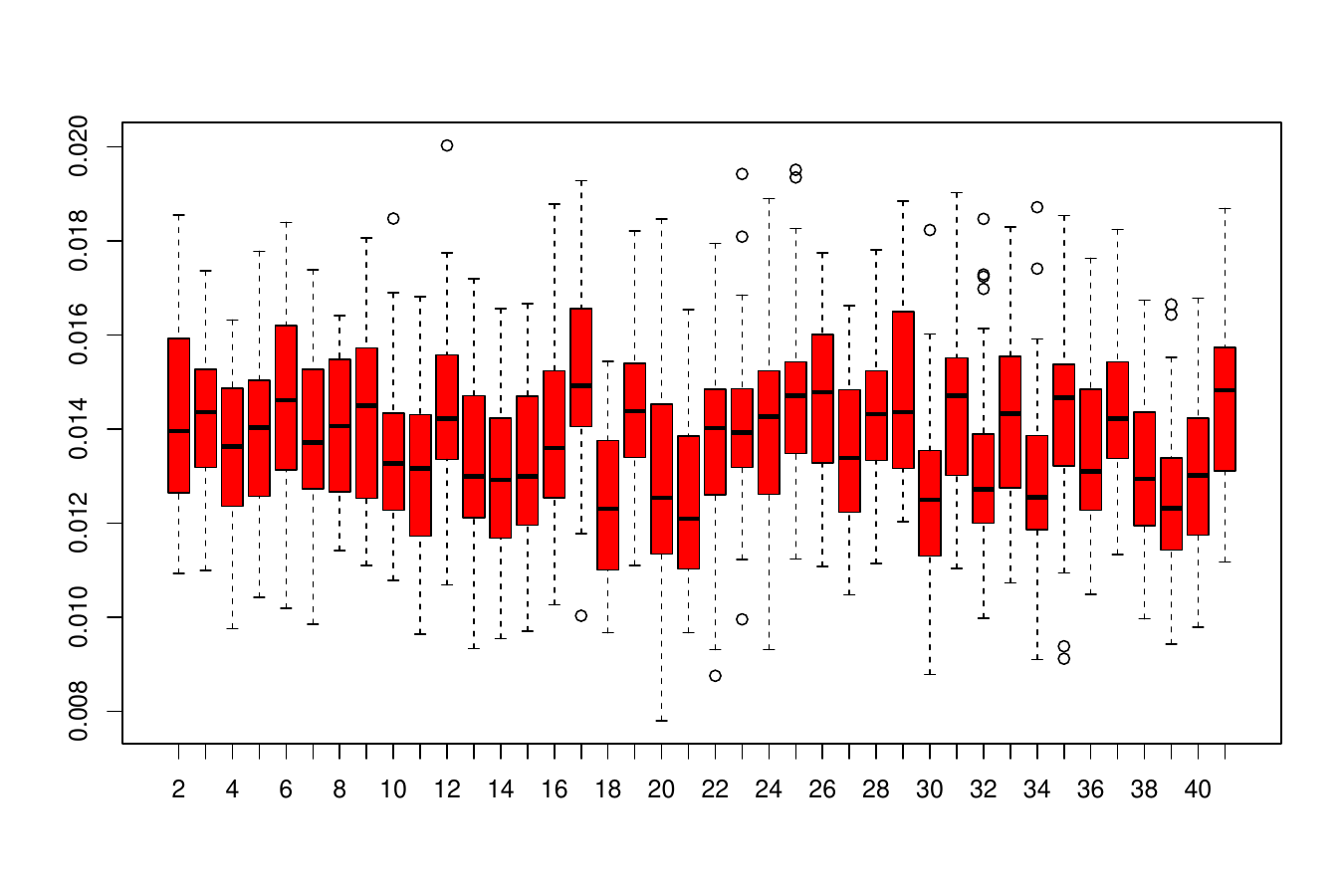}
        \caption{Poisson regression CI width box plot}
        \label{fig:pois_boxplot}
    \end{subfigure}
    \caption{}
    \label{fig:CI_width}
\end{figure}

Here we provide two examples of the methodology as mentioned earlier. We generate $n=2000$ samples for Poisson and negative binomial regression with $\log$ link. The model is described in \ref{poisson_nb_regression}. In each setup, we have a $\bf{X}_{n\times p}$ design matrix, where $p=41$. The first column of the design matrix is $\bf{1}$. The true coefficient vector $\betavec:=\left(\beta_{0},\beta_{1},\cdots,\beta_{p}\right)$ is defined as 
\[
\beta_{i}=
\begin{cases}
.5 & \text{if } i=0 \\
\frac{i}{15} & \text{if } 1\leq i\leq 10\\
0  & \text{if } 11\leq i\leq p.
\end{cases}
\]
First, the rows of the design matrix are generated from a multivariate Normal distribution with a mean vector where the elements generated uniformly from $(-2,2)$ and an identity covariance matrix. Secondly, we generate a true response variable from Poisson and Negative Binomial with mean parameter $\exp(\bf{X}\betavec)$ and for Negative Binomial we fix the dispersion parameter as $\theta=4.5$. Now we perform the Algorithm \ref{alg:plr_algo} with $B=50$ and we experiment 50 times. Now for each of the experiments either the true coefficient falls inside the calculated confidence interval or it does not. Among all the experiments we calculate how many times it does include the true value coefficient for all $\beta_{i}, i>0$. The confidence interval (CI) rate is defined as the proportion of intervals 
that contain the true parameter value, i.e., 
\[\text{CI Rate} = \frac{\text{Number of intervals containing true value}}{\text{Total intervals computed}}.\] It is challenging to address this problem because the Tweedie and Generalized Linear Model (GLM) frameworks introduce significant non-linearity. The objective is not solely to achieve a high confidence interval (CI) coverage rate, but also to ensure that the size or length of the confidence interval remains reasonable. Striking a balance between accurate coverage and a compact interval length is critical for reliable and interpretable inference.
For the nonzero coefficients, the confidence rates are presented in the table \ref{tab:CI_rate}. For the true non-zero coefficients, the average CI rate is .97 and .89 for Poisson regression and Negative Binomial regression respectively. The average CI rate for the true zero coefficients is .778 for the Poisson case and .786 for the Negative Binomial case. The confidence interval width is plotted as a box plot in the figure \ref{fig:CI_width}.
\begin{table}[ht]
\centering
\begin{tabular}{rrrr}
  \hline
 & true beta & NB CI rate & Pois CI rate \\ 
  \hline
1 & 0.50 & 0.96 & 1.00 \\ 
  2 & 0.07 & 1.00 & 1.00 \\ 
  3 & 0.13 & 1.00 & 0.98 \\ 
  4 & 0.20 & 0.98 & 1.00 \\ 
  5 & 0.27 & 0.94 & 1.00 \\ 
  6 & 0.33 & 0.96 & 1.00 \\ 
  7 & 0.40 & 0.66 & 1.00 \\ 
  8 & 0.47 & 1.00 & 1.00 \\ 
  9 & 0.53 & 1.00 & 1.00 \\ 
  10 & 0.60 & 1.00 & 1.00 \\ 
  11 & 0.67 & 1.00 & 0.70 \\ 
   \hline
\end{tabular}
\caption{Confidence rate result comparison for nonzero coefficients} 
\label{tab:CI_rate}
\end{table}

\subsection{Comparison with other bootstrap methods}
In this section, we compare the Partial Lasso and Ridge Algorithm described in Algorithm~\ref{alg:plr_algo} with a couple of state-of-the-art bootstrap methods such as a Pearson residual Bootstrap with $l_{1}$ penalty described in Algorithm~\ref{alg:lasso_bootstrap_glm} and Paired Bootstrap for GLM with $l_{1}$ penalty over coefficients. For the Paired Bootstrap, for each of the bootstrap steps, we take a bootstrap sample from the data points and calculate the coefficient with $l_{1}$ penalty. After obtaining the estimates of the coefficients from each of the bootstrap sample, we calculate the 97.5\% and 2.5\% quantiles for each of the coefficients and calculate the CI. The data generation method can be found in section~\ref{Synthetic data}. The comparison results for true nonzero coefficients for Poisson and Negative-Binomial regression can be found in table~\ref{pois_comparizon} and table~\ref{nb_comparizon}. We can see that for most of the non-zero coefficients, Algorithm~\ref{alg:plr_algo} provides shorter CI in comparison with respect to the two other methods. The Algorithm~\ref{alg:lasso_bootstrap_glm} performs better than the standard paired Bootstrap method.

\begin{table}[h!]
\centering
\begin{tabular}{rrrr}
  \hline
  PLR, Algorithm~\ref{alg:plr_algo} & Algorithm~\ref{alg:lasso_bootstrap_glm} & Paired Bootstrap  \\ 
  \hline
 0.014 & 0.046 & 0.050 \\ 
 0.014 & 0.050 & 0.056 \\ 
 0.013 & 0.049 & 0.051 \\ 
 0.014 & 0.053 & 0.057 \\ 
 0.015 & 0.055 & 0.051 \\ 
 0.014 & 0.049 & 0.050 \\ 
 0.014 & 0.054 & 0.056 \\ 
 0.014 & 0.053 & 0.055 \\ 
0.013 & 0.056 & 0.057 \\ 
0.013 & 0.052 & 0.056 \\ 
   \hline
\end{tabular}
\caption{Confidence interval width comparison for nonzero coefficients for Poisson}
\label{pois_comparizon}
\end{table}

\begin{table}[h!]
\centering
\begin{tabular}{rrrr}
  \hline
  PLR, Algorithm~\ref{alg:plr_algo} & Algorithm~\ref{alg:lasso_bootstrap_glm} & Paired Bootstrap  \\ 
  \hline
0.036 & 0.047 & 0.045 \\ 
0.036 & 0.044 & 0.043 \\ 
0.038 & 0.046 & 0.043 \\ 
0.036 & 0.043 & 0.046 \\ 
0.036 & 0.043 & 0.045 \\ 
0.036 & 0.043 & 0.046 \\ 
0.036 & 0.043 & 0.041 \\ 
0.037 & 0.042 & 0.046 \\ 
 0.034 & 0.042 & 0.042 \\ 
 0.037 & 0.040 & 0.048 \\ 
   \hline
\end{tabular}
\caption{Confidence interval width comparison for nonzero coefficients for Negative-Binomial}
\label{nb_comparizon}
\end{table}

\subsection{Application on Auto Claim data with Tweedie GBM}
\begin{table}[ht]
\centering
\begin{minipage}[b]{0.4\textwidth}
    \centering
    \begin{tabular}{|l|c|l|}
    \hline
    \textbf{Variable} & \textbf{Code} & \textbf{Level} \\
    \hline
    CAR\_USE & 1 & Private \\
            & 2 & Commercial \\
    \hline
    CAR\_TYPE & 1 & Panel Truck \\
              & 2 & Pickup \\
              & 3 & Sedan \\
              & 4 & Sports Car \\
              & 5 & SUV \\
              & 6 & Van \\
    \hline
    RED\_CAR & 1 & No \\
             & 2 & Yes \\
    \hline
    REVOLKED & 1 & No \\
             & 2 & Yes \\
    \hline
    GENDER   & 1 & F \\
             & 2 & M \\
    \hline
     MARRIED  & 1 & No \\
             & 2 & Yes \\
    \hline
    \end{tabular}
    \caption{Car Usage, Type, Red Car, Revolked, and Gender}
\end{minipage}%
\hspace{1cm}
\begin{minipage}[b]{0.4\textwidth}
    \centering
    \begin{tabular}{|l|c|l|}
    \hline
    \textbf{Variable} & \textbf{Code} & \textbf{Level} \\
    \hline
   
    PARENT1  & 1 & No \\
             & 2 & Yes \\
    \hline
    JOBCLASS & 1 & Unknown \\
             & 2 & Blue Collar \\
             & 3 & Clerical \\
             & 4 & Doctor \\
             & 5 & Home Maker \\
             & 6 & Lawyer \\
             & 7 & Manager \\
             & 8 & Professional \\
             & 9 & Student \\
    \hline
    MAX\_EDUC & 1 & < High School \\
              & 2 & Bachelors \\
              & 3 & High School \\
              & 4 & Masters \\
              & 5 & PhD \\
    \hline
    AREA      & 1 & Rural \\
              & 2 & Urban \\
    \hline
    \end{tabular}
    \caption{ Marital Status, Parent Status, Job Class, Education, and Area}
    \label{tab:encoding}
\end{minipage}
\end{table}
Tweedie Gradient Boosting Machine (GBM) refers to a type of Gradient Boosting Machine that uses the Tweedie distribution as the loss function in the model. This is particularly useful when modeling data that has a combination of continuous and discrete components, such as insurance claims, where the response variable might include a mix of zero values and positive continuous values. We consider the \textbf{Auto Claim} data from the \texttt{cplm} R package (see \cite{zhang2013cplm}). This
auto insurance data set was retrieved from the SAS Enterprise Miner database. It contains 10296 records and 28 features. Table~\ref{tab:autoclaim-data} shows the description of these variables. We consider the model described in \ref{tweedie}.

\begin{figure}[h!] 
    \centering
    \includegraphics[width=.95\linewidth]{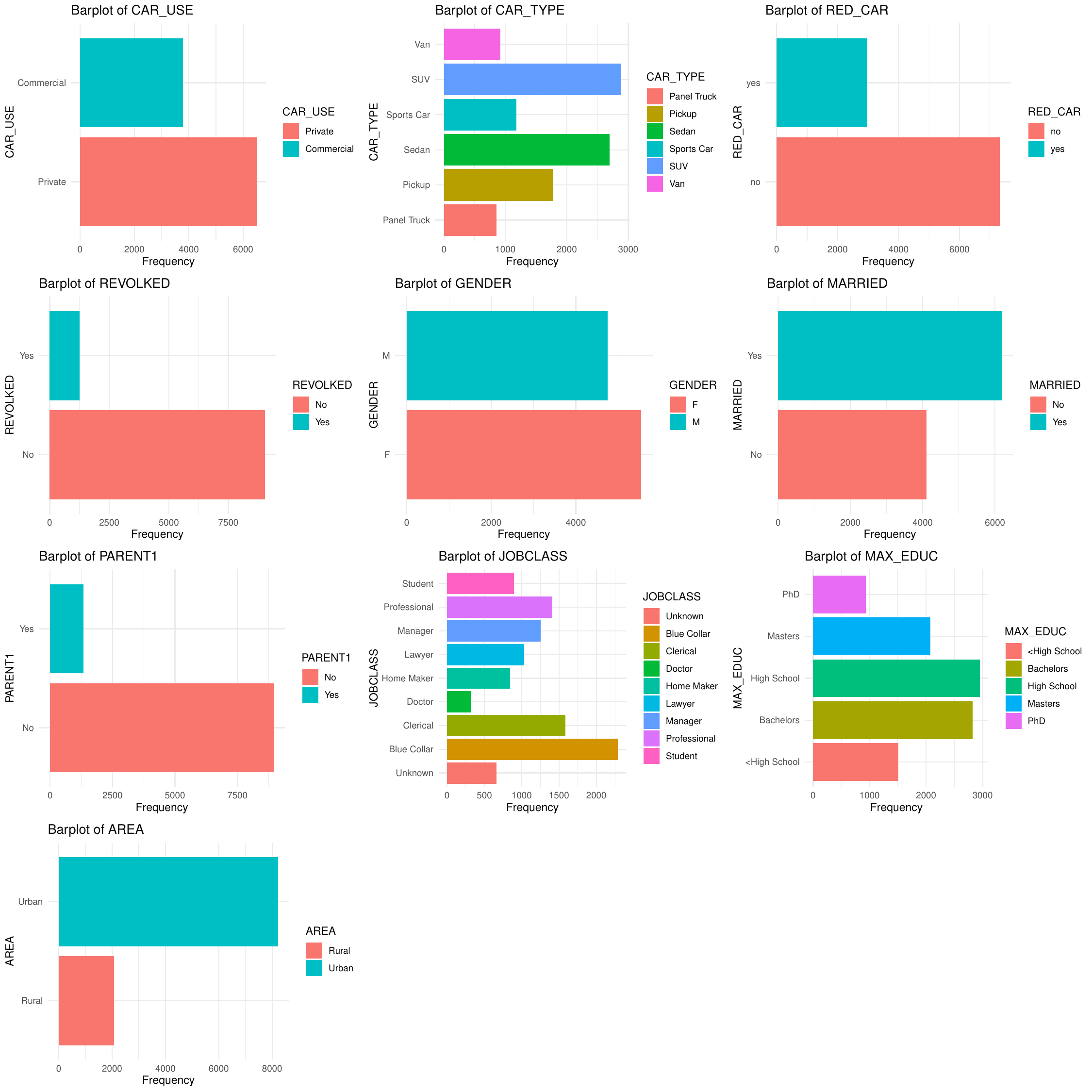} 
    \caption{Bar plots of different discrete columns in Auto claim data }
    \label{fig:barplot_discrete} 
\end{figure}

For LightGBM, we can see in the Figure~\ref{fig:barplot_discrete} that there are several categorical and continuous variables. So it is important to do a hot encoding for the categorical variables described in Table~\ref{tab:encoding}. After that one may directly use the numerical integer coded values as input for lightGBM. If some of the variables have missing values, we replace them by the median of the covariates.
\begin{figure}[h!] 
    \centering
    \includegraphics[height=0.4\linewidth,width=.4\linewidth]{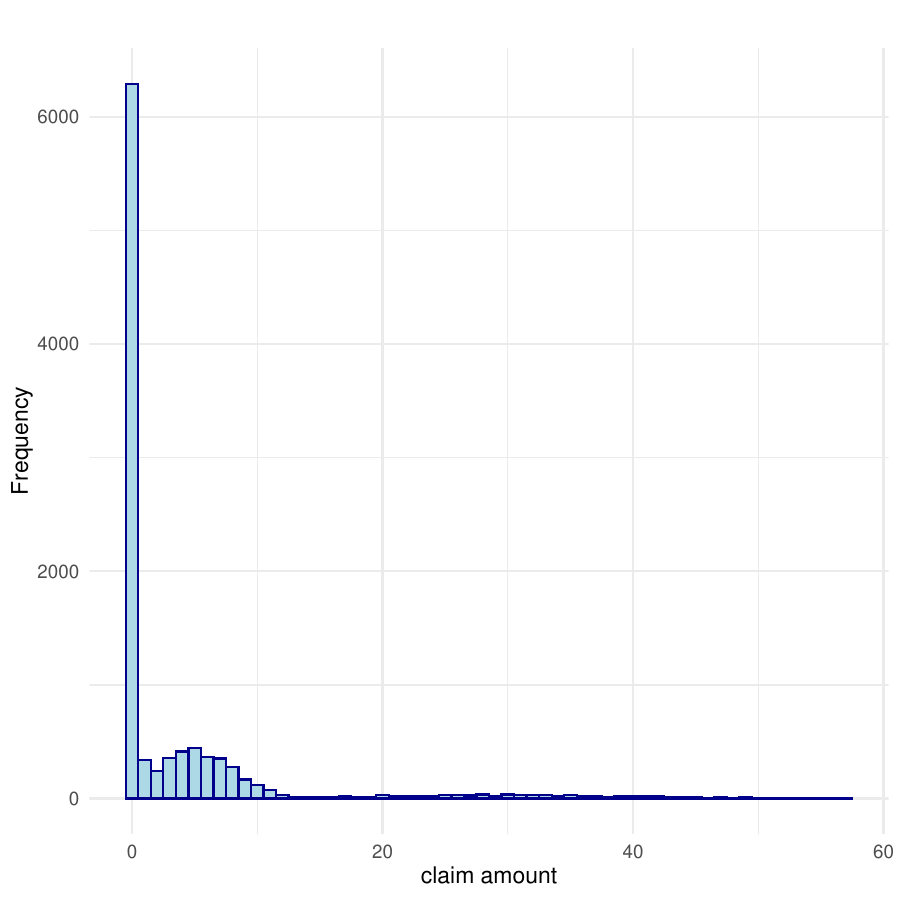} 
    \caption{Claim amount distribution with a spike at zero}
    \label{fig:claim_amount} 
\end{figure}
\begin{table}[!htbp]
\addtolength{\tabcolsep}{-0.1em}
\begin{center}\tiny
\begin{tabular}{lll}
\toprule
Variable & Type & Description\tabularnewline
\midrule
\texttt{POLICYNO} & character  & the policy number\tabularnewline
\texttt{PLCYDATE} & date  & policy effective date\tabularnewline
\texttt{CLM\_FREQ5} & integer  & the number of claims in the past 5 years\tabularnewline
\texttt{CLM\_AMT5} & integer  & the total claim amount in the past 5 years\tabularnewline
\texttt{CLM\_AMT} & integer  & the claim amount in the current insured period\tabularnewline
\texttt{KIDSDRIV} & integer  & the number of driving children\tabularnewline
\texttt{TRAVTIME} & integer  & the distance to work\tabularnewline
\texttt{CAR\_USE} & factor  & the primary use of the vehicle: ``Commercial'', ``Private''\tabularnewline
\texttt{BLUEBOOK} & integer  & the value of the vehicle\tabularnewline
\texttt{RETAINED} & integer  & the number of years as a customer\tabularnewline
\texttt{NPOLICY} & integer  & the number of policies\tabularnewline
\texttt{CAR\_TYPE} & factor  & the type of the car: ``Panel Truck'', ``Pickup'', ``Sedan'',
``Sports Car'', ``SUV'', ``Van''\tabularnewline
\texttt{RED\_CAR} & factor  & whether the color of the car is red: ``no'', ``yes''\tabularnewline
\texttt{REVOLKED} & factor  & whether the driver's license was invoked in the past 7 years: ``No'',
``Yes''\tabularnewline
\texttt{MVR\_PTS} & integer  & MVR violation records\tabularnewline
\texttt{CLM\_FLAG} & factor  & whether a claim is reported: ``No'', ``Yes''\tabularnewline
\texttt{AGE} & integer  & the age of the driver\tabularnewline
\texttt{HOMEKIDS} & integer  & the number of children\tabularnewline
\texttt{YOJ} & integer  & years at current job\tabularnewline
\texttt{INCOME} & integer  & annual income\tabularnewline
\texttt{GENDER} & factor  & the gender of the driver: ``F'', ``M''\tabularnewline
\texttt{MARRIED} & factor  & married or not: ``No'', ``Yes''\tabularnewline
\texttt{PARENT1} & factor  & single parent: ``No'', ``Yes''\tabularnewline
\texttt{JOBCLASS} & factor & ``Unknown'', ``Blue Collar'', ``Clerical'', ``Doctor'', ``Home
Maker'', ``Lawyer'', ``Manager'', ``Professional'', ``Student''\tabularnewline
\texttt{MAX\_EDUC} & factor  & max education level: ``<High School'', ``Bachelors'', ``High
School'', ``Masters'', ``PhD''\tabularnewline
\texttt{HOME\_VAL} & integer  & the value of the insured's home\tabularnewline
\texttt{SAMEHOME} & integer  & years in the current address\tabularnewline
\texttt{DENSITY} & factor  & home/work area: ``Highly Rural'', ``Highly Urban'', ``Rural'',
``Urban''\tabularnewline
\bottomrule
\end{tabular}
\par\end{center}
\caption{Description of variables in the \textbf{Auto Claim} data}
\label{tab:autoclaim-data}
\end{table}
We take \texttt{CLM\_AMT5} as the dependent variable (\(Y\)), and all the other
variables except \texttt{POLICYNO}, \texttt{PLCYDATE}, \texttt{CLM\_FREQ5} and
\texttt{CLM\_AMT}, as the independent variables (\(\bX\)). We can see the distribution of the number and amount of claims in figure \ref{fig:claim_amount} where many people had zero claims, for which there is a spike in zero. We plot the discrete variables' bar plot in the figure \ref{fig:barplot_discrete}. Now using the Algorithm we can see that in table \ref{tab:ci_tweedie}, MVR violation records, CAR TYPE (pickup and sedan),
whether the driver’s license was invoked in the past 7 years (Yes), Married (yes), Jobclass (Doctor and lawyer), Area (urban) might be key individual factors to determine the higher claim amount with a positive confidence interval away from zero. 

\begin{table}[ht]
\centering
\resizebox{.5\textwidth}{!}{
\begin{tabular}{rlrrr}

  \hline
 & variable & lower\_CI & upper\_CI & length\_of\_CI \\ 
  \hline
1 & intercept & -0.441 & -0.125 & 0.316 \\ 
  2 & KIDSDRIV & 0.000 & 0.054 & 0.054 \\ 
  3 & TRAVTIME & -0.000 & 0.000 & 0.000 \\ 
  4 & BLUEBOOK & -0.000 & 0.000 & 0.000 \\ 
  5 & NPOLICY & -0.000 & 0.000 & 0.000 \\ 
  6 & MVR\_PTS & 0.175 & 0.199 & 0.024 \\ 
  7 & AGE & -0.000 & 0.000 & 0.000 \\ 
  8 & HOMEKIDS & -0.000 & 0.000 & 0.000 \\ 
  9 & YOJ & -0.000 & 0.000 & 0.000 \\ 
  10 & INCOME & -0.000 & -0.000 & 0.000 \\ 
  11 & HOME\_VAL & -0.000 & -0.000 & 0.000 \\ 
  12 & SAMEHOME & -0.000 & 0.000 & 0.000 \\ 
  13 & CAR\_USE\_Commercial & 0.000 & 0.000 & 0.000 \\ 
  14 & CAR\_TYPE\_Pickup & -0.072 & -0.000 & 0.072 \\ 
  15 & CAR\_TYPE\_Sedan & -0.237 & -0.000 & 0.237 \\ 
  16 & CAR\_TYPE\_Sports.Car & 0.000 & 0.255 & 0.255 \\ 
  17 & CAR\_TYPE\_SUV & -0.000 & 0.000 & 0.000 \\ 
  18 & CAR\_TYPE\_Van & -0.000 & 0.167 & 0.167 \\ 
  19 & RED\_CAR\_yes & -0.000 & 0.000 & 0.000 \\ 
  20 & REVOLKED\_Yes & 1.454 & 1.604 & 0.150 \\ 
  21 & GENDER\_M & -0.000 & 0.000 & 0.000 \\ 
  22 & MARRIED\_Yes & -0.119 & 0.000 & 0.119 \\ 
  23 & PARENT1\_Yes & -0.000 & 0.000 & 0.000 \\ 
  24 & JOBCLASS\_Blue.Collar & -0.000 & 0.000 & 0.000 \\ 
  25 & JOBCLASS\_Clerical & -0.000 & 0.000 & 0.000 \\ 
  26 & JOBCLASS\_Doctor & -0.458 & -0.000 & 0.458 \\ 
  27 & JOBCLASS\_Home.Maker & -0.000 & 0.000 & 0.000 \\ 
  28 & JOBCLASS\_Lawyer & -0.338 & -0.000 & 0.338 \\ 
  29 & JOBCLASS\_Manager & -0.000 & 0.000 & 0.000 \\ 
  30 & JOBCLASS\_Professional & -0.000 & 0.000 & 0.000 \\ 
  31 & JOBCLASS\_Student & -0.000 & 0.000 & 0.000 \\ 
  32 & MAX\_EDUC\_Bachelors & -0.001 & 0.000 & 0.001 \\ 
  33 & MAX\_EDUC\_High.School & -0.000 & 0.000 & 0.000 \\ 
  34 & MAX\_EDUC\_Masters & -0.000 & 0.000 & 0.000 \\ 
  35 & MAX\_EDUC\_PhD & -0.000 & 0.000 & 0.000 \\ 
  36 & AREA\_Urban & 0.972 & 1.245 & 0.273 \\ 
   \hline
\end{tabular}
}
\caption{Confidence Intervals for Auto-Claim Variables} 
\label{tab:ci_tweedie}
\end{table}

\section{Conclusion} In this article, we have described several methodologies for selective inference that can be used for the estimation of the confidence interval of important features in GLM. Based on our experiment in synthetic data and real data application, Algorithm $\ref{alg:plr_algo}$ is a potential alternative method for estimating the confidence interval of important coefficients in Tweedie regression and other GLM problems with high confidence rate and shorter confidence interval. Extensions to the distributed machines and theoretical developments are kept for future study. We also developed a new method for distribution free inference with conformal analysis for prediction interval estimation for LightGBM and GLM with Tweedie error in \cite{manna2025distribution}. The codes for our simulations are kept in \url{https://github.com/alokesh17/selective_inference-}.

\section{Author Credit statement} In this paper, Alokesh Manna (\url{https://sites.google.com/view/alokesh-manna/home} and \url{https://statistics.uconn.edu/person/alokesh-manna/}) served as the lead author, contributing to the writing, conceptualization, methodology, validation, formal analysis, and investigation. Dr. Zijian Huang is a Quantitative Analytics Specialist at Wells Fargo (\url{https://www.linkedin.com/in/zijian-huang-60a2076b/}) who helped with getting different literature surveys, computation, and Tweedie formulation for this paper. Dr. Yuwen Gu (\url{https://statistics.uconn.edu/yuwen-gu/}) helped conceptualize formally incorporate it into the LightGBM framework. Dr. Dipak Dey (\url{https://statistics.uconn.edu/person/dipak-dey/}) motivated us to compare the different methods and write the overall paper with his thoughtful suggestions.  Robin He is Sr. Director \& Data Scientist at Travelers, (\url{https://www.linkedin.com/in/jichaohe/}), contributed to problem formulation across various real-world scenarios and practical applications of our approach.

\clearpage
\bibliographystyle{apalike}
\bibliography{lgb1}
\end{document}